\def\templateType{article}
\def\tTwiley{wiley}
\def\tTarticle{article}

\ifx\templateType\tTwiley
\documentclass[alpha-refs]{wiley-article}
\fi
\ifx\templateType\tTarticle
\documentclass[]{article}
\fi

\usepackage[utf8]{inputenc}
\usepackage[french,english]{babel}
\usepackage{url}
\usepackage{graphicx}
\usepackage{multicol, blindtext}
\usepackage{multirow}
\usepackage{bm}
\usepackage{placeins} % pour \FloatBarrier
\usepackage{array}
\usepackage{amsmath,amssymb}
\usepackage{bbold}
\usepackage{dsfont}
\usepackage{color}
\usepackage[dvipsnames]{xcolor}
\usepackage{appendix}
\let\oldappendices\appendices
\def\appendices{\oldappendices\adjustmtc}
\usepackage[linesnumbered,ruled,vlined]{algorithm2e}
\usepackage{chngcntr}
\usepackage{etoolbox}
\usepackage{longtable}
\setlongtables
\usepackage{caption}
\usepackage{subfig}
\usepackage{textcase}
\usepackage{parskip}
\usepackage[normalem]{ulem}
\usepackage{soul}
\usepackage{siunitx}
\usepackage{authblk}
\usepackage{natbib}
\usepackage{colortbl}
\usepackage{hhline}
\usepackage{booktabs}
\setlongtables

\DeclareMathOperator*{\argmin}{argmin}

\newcommand{\ome}{\omega_{e}}
\newcommand{\omep}{\omega_{e^\prime}}
\newcommand{\pe}{p_{e}^{m_e}}
\newcommand{\pep}{p_{e^\prime}^{m_{e^\prime}}}
\newcommand{\xe}{x_{e}^{m_e}}
\newcommand{\xep}{x_{e^\prime}^{m_{e^\prime}}}
\newcommand{\me}{m_{e}}
\newcommand{\mep}{m_{e^\prime}}
\newcommand{\Me}{M_{e}}
\newcommand{\Mep}{M_{e^\prime}}

\newcommand{\omeps}{\omega_{e^\prime;s}}
\newcommand{\xes}{x_{e;s}^{m_e}}
\newcommand{\xeps}{x_{e^\prime;s}^{m_{e^\prime}}}
\newcommand\red[1]{#1}
\definecolor{Lightgray}{gray}{0.8}
\usepackage{array}
\newcommand{\PreserveBackslash}[1]{\let\temp=\\#1\let\\=\temp}
\newcolumntype{C}[1]{>{\PreserveBackslash\centering}p{#1}}
\newcolumntype{R}[1]{>{\PreserveBackslash\raggedleft}p{#1}}
\newcolumntype{L}[1]{>{\PreserveBackslash\raggedright}p{#1}}

\title{Sequential Aggregation of Probabilistic Forecasts - Applicaton to Wind Speed Ensemble Forecasts}
\author[1,3]{Micha\"el Zamo}
\author[2]{Liliane Bel}
\author[1,3]{Olivier Mestre}

\affil[1]{M\'et\'eo-France, Direction des Op\'erations pour la Production, 42 avenue Gaspard Coriolis, 31057 Toulouse cedex 07, France.}
\affil[2]{Universit\'e Paris-Saclay, AgroParisTech, INRAE, UMR MIA-Paris, 75005, Paris, France}
\affil[3]{CNRM/GAME, M\'et\'eo-France/CNRS URA 1357, Toulouse, France}

\ifx\templateType\tTwiley
\papertype{Original Article}
\contrib[\authfn{1}]{Equally contributing authors.}
\corraddress{Micha\"el Zamo, M\'et\'eo-France, Direction des Op\'erations pour la Production, 42 avenue Gaspard Coriolis, 31057 Toulouse cedex 07, France.}
\corremail{michael.zamo@meteo.fr}
\runningauthor{Micha\"el Zamo et al.}
\fi

\begin{document}

\maketitle

\begin{abstract}
\red{In the field of numerical weather prediction (NWP), the
  probabilistic distribution of the future state of the atmosphere is
  sampled with Monte-Carlo-like simulations, called ensembles. These
  ensembles have deficiencies (such as conditional biases) that can be
  corrected thanks to statistical post-processing methods. Several ensembles exist and
  may be corrected with different statistiscal methods. A further step
  is to combine these raw or post-processed ensembles. The theory of
  prediction with expert advice allows us to build combination algorithms
  with theoretical guarantees on the forecast performance. This
  article adapts this theory to the case of probabilistic forecasts
  issued as step-wise cumulative distribution functions (CDF). The
  theory is applied to wind speed forecasting, by combining several
  raw or post-processed ensembles, considered as CDFs. The second goal
  of this study is to explore the use of two forecast performance
  criteria: the Continous ranked probability score (CRPS) and the
  Jolliffe-Primo test. Comparing the results obtained with both
  criteria leads to reconsidering the usual way to build skillful
  probabilistic forecasts, based on the minimization of the
  CRPS. Minimizing the CRPS does not necessarily produce reliable
  forecasts according to the Jolliffe-Primo test. The Jolliffe-Primo
  test generally selects reliable forecasts, but could lead to issuing
  suboptimal forecasts in terms of CRPS. It is proposed to use both criterion to
  achieve reliable and skillful probabilistic forecasts.} 
% Please include a maximum of seven keywords
%\keywords{keyword 1, \emph{keyword 2}, keyword 3, keyword 4, keyword 5, keyword 6, keyword 7}
\end{abstract}

\section{Introduction}

As a chaotic dynamical system, the atmosphere has an evolution that is
intrinsically uncertain
\citep{malardel2005fondamentaux,holton2012introduction} \red{and should be described in a probabilistic form}. In the field
of numerical weather prediction (NWP), \red{this probabilistic form is not a probability distribution but a set of deterministic forecasts whose aim is to assess the forecast uncertainty
\citep{leutbecher2008ensemble}. Such a set of deterministic forecasts is called an ensemble forecast and each individual deterministic forecast is called a member}. The members are usually
obtained by running the same NWP model with different initial
conditions and different parametrizations of the model physics
\citep{descamps2011representing}. Forecast uncertainty can then be derived
from the members as a probability distribution with statistical
estimation techniques and considering the members are a random sample from an unknown multivariate probability distribution .

Being often biased and under-dispersed for
surface parameters \citep{hamill1998evaluation,buizza2005comparison},
the ensemble forecast systems may be post-processed with statistical
methods, called ensemble model output statistics (EMOS) to get more
skillful forecast distributions
\citep{wilson2007calibrated,thorarinsdottir2010probabilistic,moller2013postprocessing,zamo2014benchmarkII,baran2015log,taillardat2016calibrated}.

Nowadays, several ensemble forecast systems are available routinely
\citep{bougeault2010thorpex,descamps2014pearp}. Combining, or
``aggregating'', several forecasts may improve the predictive
performance compared to the most skillful post-processed ensemble
\citep{allard2012probability,gneiting2013combining,baudin2015prevision,baran2016mixture,moller2016probabilistic,bogner2017combining}.
The theory of prediction with expert advice
\citep{cesa2006prediction,stoltz2010agregation} shows how to
efficiently aggregate in real-time several forecasts based on their
respective past performances. This theory studies the mathematical
properties of aggregation algorithms of several forecasting systems (called
``experts'' in this framework), and has been applied mostly to point forecasts.
%whether for point forecasts \citep{fritsch2000model,baars2005performance,woodcock2005operational} or probabilistic forecasts

\red{The first goal of the present work is to apply the
  theory of  prediction with expert advice to probabilistic forecasts
  represented as step-wise cumulative distribution functions (CDF)
  with any number of steps. Two previous studies
  \citep{baudin2015prevision,thorey2017thesis} used this
  theory for specific cases of probabilistic forecasts. In
  \cite{baudin2015prevision}, the experts are the \emph{ordered}
  individual members of pooled ensembles. Each expert's forecast is a
  stepwise cumulative distribution function with one step. In this case the experts are not
  identifiable over time although it is required by the theory. For instance, at
  different times, the lowest forecast value comes from a different
  member of a different ensemble. \cite{thorey2017thesis}
  applies the theory of prediction with expert advice to forecasts of
  photovoltaic electricity production. The experts are ensemble
  forecasts or built from ensemble or deterministic forecasts thanks
  to statistical regression methods. Each expert is treated as a
  different deterministic forecast, even though some of them are
  members of the same ensemble. Other experts are built thanks to
  quantile regression methods. As such, the set of forecast quantiles
  could be considered as a specific probabilistic expert, but each
  forecast quantile is again considered as a separate deterministic
  expert.} In this study, contrary to \cite{baudin2015prevision} and
\cite{thorey2017thesis}, each expert is a whole (raw or
post-processed) ensemble, and is thus actually identifiable over
time. \red{This work extends the work of \cite{baudin2015prevision} in
  the sense that the formulae established in this article reduce to
  the ones in \cite{baudin2015prevision} if considering step-wise CDFs
  with a single step. } In a nutshell, the aggregated forecast is a linear
combination of the CDFs of the experts. Since the aggregated forecast
must be a CDF, only convex aggregation strategies are investigated:
the weights are constrained to be positive and to sum up to one. 

\red{The second goal of this work is to compare two model selection approaches when dealing with probabilistic forecasting systems \citep{collet2017generic}. The first approach is based on ``reliability'' (or ``calibration'') and ``sharpness'' \citep{gneiting2007probabilistic,jolliffe2011fverification}. A forecasting system is reliable if the conditional probability distribution
  of the observation given the forecast distribution is equal to the forecast distribution. A forecasting system is sharper when, on average, it predicts a lower dispersion of the observation. According to the \emph{sharpness-calibration paradigm} of \cite{gneiting2007probabilistic} a forecasting system should aim at providing reliable probabilistic
  forecasts that are the sharpest (i.e. less dispersed) possible. A
  practical motivation of this sharpness-calibration paradigm is that
  decisions based on such a forecasting system would be optimal due
  to the reliability of the forecast and less uncertain due to the
  forecast's low dispersion, which improves its value for economical decisions
  \citep{richardson2001measures,zhu2002economic,mylne2002decision}. Among several reliable forecasting systems, one should thus select the sharpest. The second approach to model selection among probabilistic forecasting systems is based on a scoring rule, such as the Continuous Ranked
  Probabilistic Score (CRPS, \citeauthor{matheson1976scoring}
  \citeyear{matheson1976scoring}): the selected model is the one that has the best value of the scoring rule (highest or lowest value depending on the scoring rule). The two approaches to model selection do not yield equivalent forecasts, as previously mentioned in different
  studies \citep{collet2017generic,wilks2018enforcing}. For instance, minimizing the
  CRPS may lead to forecasts that are not reliable. To solve this problem,
  \cite{wilks2018enforcing} proposes to minimize the CRPS penalized
  with a term quantifying the unreliability of the
  forecast. \cite{collet2017generic} introduces a post-processing method
  that, under quite strong assumptions, yields reliable forecasts
  without degrading too much the CRPS compared to the CRPS-minimizing
  approach. We do not use any of these solutions here but compare the two model selection approaches and their properties on a case study. Both approaches are used to select the best forecasting system among several experts and aggregated forecasts. In the first approach, reliability is imposed by using the Jolliffe-Primo flatness test (JP test,  \cite{jolliffe2008evaluating}). As for the second approach, forecast performance is measured with the CRPS.}
% The desired properties of the aggregation are two-fold. The first one
% is to yield better forecast performances. This motivates using the theory
% of prediction with expert advice to ensure good forecast
% performance. The second goal is to dynamically tackle changes in the
% ensemble models, that may strongly affect the performance of the raw
% or post-processed ensembles. A good aggregation method should quickly
% detect changes in the performance of the experts and adapt the
% aggregation weights to discard the bad ones and favor the good
% ones.

% In this work, aggregation is applied to forecasts of the 10~m wind speed over
% France. Forecasting surface wind is quite difficult due to complex
% interactions of phenomenon at different spatial scales. For
% instance, wind speed is influenced by large scale atmospheric
% structures such as cyclones and anticyclones, but also by local
% orography and surface friction. Local atmospheric effects such as downward drifts under convective
% clouds also contribute to the direction and speed of wind. National
% weather services need to have skillful wind speed forecasts to
% issue early warnings to the population and civil security
% services. Also, wind speed influences many economic activities, such
% as sailing, windpower generation, construction,... that need good
% forecasts for decision making.

In Section~\ref{sec:aggr_aggr}, the theory of prediction with expert
advice is presented, along with notations. It is
shown how this theory can be straightforwardly applied to step-wise
CDFs. The CRPS and JP test are also introduced with more
details and their use is further motivated. Section~\ref{sec:aggr_weights} presents the different
aggregation methods investigated. Some are empirical, while others
exhibit interesting theoretical
properties. Section~\ref{sec:aggr_experts} describes the use-case of
this study and the data it uses: four ensemble forecasts, two EMOS methods used to
post-process the ensembles, and the wind speed observation. The
results of the comparison of the aggregation methods are presented in
Section~\ref{sec:aggr_results}. These results motivate a more
theoretical comparison of the two approaches to model selection among
probabilistic forecasts, in Section~\ref{sec:aggr_discussion}. Finally,
Section~\ref{sec:aggr_conclusion} concludes with a summary of the
results and perspectives.

\section{Theoretical Framework and Performance Assessment Tools}
\label{sec:aggr_aggr}

The desired properties of forecast aggregation are two-fold. The first one
is to yield an aggregated forecast that performs better than any of the forecasts that are used in the aggregation. According to the theory of
prediction with expert advice
\citep{cesa2006prediction,stoltz2010agregation}, some algorithms used
to sequentially aggregate forecasts exhibits theoretical guarantees of
performance. These guarantees state that the aggregated forecast will
not perform much worse than some skillful reference forecast, called
the oracle. In practice, the aggregated forecast may even outperform
the oracle. This motivates using the theory of prediction with expert
advice. The second desired property is to dynamically tackle changes
in the forecasts' generating process (such as modification in NWP model's
code). These changes may strongly affect the performance of the raw or
post-processed forecasts. A good aggregation method should quickly
detect changes in the performance of the individual forecasts and
adapt the aggregation weights to discard the bad ones and favor the
good ones. Being a sequential aggregation framework based on the
recent performance of the experts, the theory of prediction with
expert advice may help in reaching this second goal. 

\subsection{Sequential Aggregation of Step-Wise CDFs}

The situation tackled by the theory of prediction with expert advice
is the following: a forecaster has to forecast some parameter of
interest by using only past observations of the parameter and past and
current forecasts of the parameter steming from several
sources. \red{These sources whose forecasts are to be aggregated are
  called ``experts'' in this theory. In this very general framework no
  assumption is made on the generating process of the observations and the experts.} 

More formally, at time $t$, before the observation $y_t \in
\mathcal{Y}$ is revealed, let us suppose available the forecast of the
$E \in \mathbb{N}^*$ experts and of the past observations
(noted $y_{j=1,\ldots,t-1}$). An expert is any means, in a very
general sense (NWP model, human expertise, \ldots), to produce a
forecast of $y_t$ at each time $t$, before the
observation $y_t$ is known. The forecast of expert $e \in
\{1,\ldots, E\}$ at time $t$ is noted $\widehat{y}_{e;t} \in
\widehat{\mathcal{Y}}$, with $\widehat{\mathcal{Y}}$ the value set of the
forecasts. \red{Although the prediction with expert advice has
 been mostly used with experts issuing point forecasts ($\widehat{\mathcal{Y}} \subseteq \mathbb{R}$), let us stress
 that the experts yield forecasts of any type, not necessarily point forecasts. The
 theory straightforwardly adapts to probabilistic forecasts as will be
 shown below. Someone or something, called the ``forecaster'' in the theory,}
produces an aggregated forecast $\widehat{y}_t$ as a linear combination of the experts'
current forecasts
\begin{equation}
  \widehat{y}_t=\sum_{e=1}^E\omega_{e;t}\widehat{y}_{e;t},
\end{equation}
where $\omega_{e;t} \in \Omega \subseteq \mathbb{R}$ is the aggregation weight of expert
$e$ at time $t$. The
aggregation weights are computed using only information available at
the present time, namely the past and present experts' forecasts
$\widehat{y}_{e;j}$ with $e=1,\ldots,E$ and $j=1,\ldots,t$, and the
past observations $y_j$, with $j=1,\ldots,t-1$. The aggregation is
usually initialized with equal weights, that is,
$\omega_{e;1}=\frac{1}{E},\quad \forall e=1,\ldots,E$. The weights
$\omega_{e;t}$ can be computed with many algorithms (called
aggregation methods), some of which are presented in
Section~\ref{sec:aggr_weights}. 

When the observation $y_t$ is revealed, the forecaster suffers a loss,
quantified with a function $\ell:\widehat{\mathcal{Y}} \times
\mathcal{Y} \rightarrow \mathbb{R}$. The most general goal of the
forecaster would be to build the best possible forecast, that is, to
minimize its cumulative loss over a period of time
$t=1,\ldots,T$. This minimization is not possible for all sequences
${(\hat y_{e,t},y_t)}_{e,t}$, since it is always theoretically
possible to build a sequence of observations that makes the
forecaster's cumulative loss arbitrarily high. Therefore, a more
realistic goal is to build the best possible forecast relatively to
the best element from some class of reference forecasts. \red{Let us note
$\mathcal{C}$ such a class of reference forecasts, whose elements are written
\begin{equation}
  \widetilde{y}_t=\sum_{e=1}^E\widetilde{\omega}_{e;t}\widehat{y}_{e;t},
\end{equation}
In this case the weights $\widetilde{\omega}_{e;t}$} may be computed by using the present and
future information (that is on the whole period $t^\prime \in \{1,\ldots,T\}$). Hence the aggregated forecast delivered by the forecaster may be different from the best aggregated forecast in class $\mathcal{C}$ (called the oracle). Two examples of class $\mathcal C$ are the set of the
available experts, or the set of linear combination of the experts'
forecasts with constant weighting computed with some chosen
aggregation method. The computation of the oracle uses all the information
for the whole period. Thus, the oracle cannot be used for real-time
applications, but may be used as a reference in order to evaluate
aggregation methods. The regret $R^\mathcal{C}_T$ of the forecaster
relatively to the class $\mathcal{C}$ is the cumulative additional
loss suffered by the forecaster who used its own aggregated forecast
instead of the oracle of class $\mathcal{C}$:
\begin{equation}
  R^\mathcal{C}_T=\sum_{t=1}^T\ell(\widehat{y}_t,y_t) -
  \inf_{\widetilde{y}_t \in \mathcal{C}}\sum_{t=1}^T\ell(\widetilde{y}_t,y_t). \label{eq:aggr_regret}
\end{equation}
According to the theory of prediction with expert advice there exists aggregation methods available to the forecaster such that the regret relative to class $\mathcal{C}$ is sub-linear in $T$
\begin{equation}
  \sup R^\mathcal{C}_T \le o(T) \label{eq:aggr_regret_sublin},
\end{equation}
where the supremum is taken over all possible sequences of
the observation and experts. \red{Let's point out that the regret is
  not lower bounded, and so may be negative for some datasets.} Since
the upper bound on the regret holds for \emph{any} sequence of
observation and forecasts, this sublinearity property ensures to the
forecaster good forecast performances. \red{In most cases,
  this property only requires that the chosen loss $\ell$ be convex in
  its first argument, and no assumption is required about the observation or the experts
  \citep{cesa2006prediction}. In specific cases where some information
  is known about the experts, such as a correlation structure between
  the experts' forecasts, this information may be used to design
  better aggregation methods (see supplement in
  \cite{swinbank2016tigge}). \cite{adjakossa2020kalman} uses assumptions about the experts' generating process to improve aggregation methods for point forecasts. Here only the general case is treated.}
\cite{mallet2007description} and \cite{gerchinovitz2008further} review
many aggregation methods of expert advice, with numerical algorithms
thereof. Both articles describe the case of experts \red{issuing point
  forecasts} ($\widehat{\mathcal{Y}} \subseteq \mathbb{R}$) and real
observations ($\mathcal{Y} \subseteq \mathbb{R}$), along with
theoretical bounds for the $L_2$-loss
($\ell(\widehat{y_t},y_t)=(\widehat{y}_t-y_t)^2$), when they exist.

\red{Applying this theory for probabilistic forecasts expressed as
  step-wise CDF is straightforward.} The observation is supposed real-valued ($\mathcal{Y} \subseteq
\mathbb{R}$). The forecast space $\widehat{\mathcal{Y}}$ is the set of
step-wise CDF, that is the set of piece-wise constant, non-decreasing
functions taking their values in $[0;1]$. Each expert forecast $\widehat{y}_{e;t}$ is a
step function with jumps of heights $p_e^{m_e}$ (called weights) at
$M_e \in \mathbb{N}^*$ values $x_{e;t}^{m_e} \in \mathcal{Y}$. For instance the values $x_{e;t}^{m_e}$ may come from an ensemble forecast with $M_e$ members. The weights are such that
$p_e^{m_e} > 0$, for $m_e=1,\ldots,M_e$, and
$\sum_{m_e=1}^{M_e} p_e^{m_e} = 1$. Then each expert's forecast is the
step-wise CDF 
\begin{equation}
\widehat{y}_{e;t}=\sum_{m_e=1}^{M_e}p_e^{m_e}H(x-x_{e;t}^{m_e})
\end{equation}
where $H$ is the Heaviside function, and $x \in \mathcal{Y}$. Without
loss of generality, the $x_{e;t}^{m_e}$ are supposed sorted in
ascending order for each expert and at each time $t$, so that
$x_{e;t}^{m_e}$ may be considered as the quantile of order $\tau_e^{m_e} =
\sum_{m^\prime_e=1}^{m_e} p_e^{m^\prime_e}$ of a random variable
$\widehat{Y}_{e;t} \in \mathcal{Y}$.

The aggregated forecast CDF, $\widehat{y}_t$, is a step function at
the pooled values $\{x_{e;t}^{m_e}; m_e=1,\ldots,M_e, e=1,\ldots,E\}$
such that the jump of height $\omega_{e;t}p_e^{m_e}$ is associated to
the value $x_{e;t}^{m_e}$, thus

\ifx\templateType\tTarticle
\begin{align}
  \widehat{y}_t(x_{e;t}^{m_e}) & = \sum_{e^\prime=1}^E\omega_{e^\prime;t}\left(\sum_{m^\prime_{e^\prime}=1}^{M_{e^\prime}}p_{e^\prime}^{m^\prime_{e^\prime}}H(x_{e;t}^{m_e} - x_{e^\prime;t}^{m^\prime_{e^\prime}})\right)  = \tau_{e;t}^{m_e}. \label{eq:aggregation_order}
\end{align} 
\else
\begin{align}
  \widehat{y}_t(x_{e;t}^{m_e}) & = \sum_{e^\prime=1}^E\omega_{e^\prime;t}\left(\sum_{m^\prime_{e^\prime}=1}^{M_{e^\prime}}p_{e^\prime}^{m^\prime_{e^\prime}}H(x_{e;t}^{m_e} - x_{e^\prime;t}^{m^\prime_{e^\prime}})\right) & = \tau_{e;t}^{m_e}. \label{eq:aggregation_order}
\end{align} 
\fi

\begin{figure}[!ht]
  \centering
  \includegraphics[width=0.8\columnwidth, angle=-90]{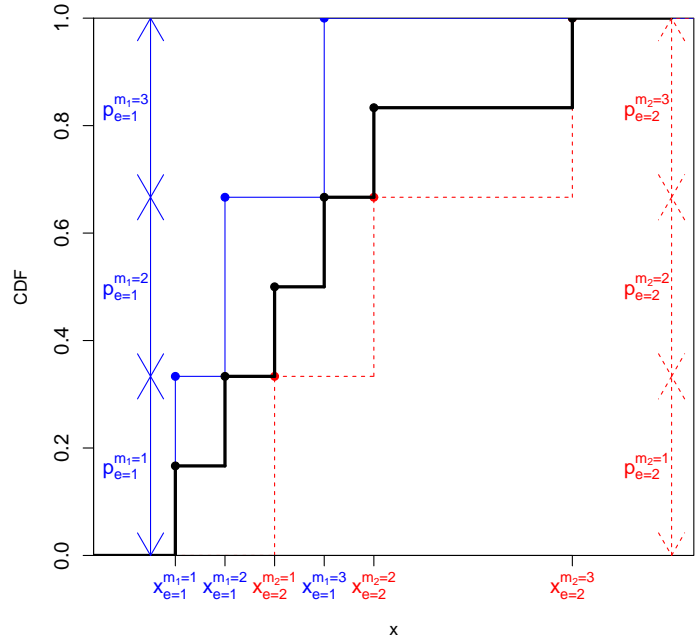}
  \caption[Example of aggregation of two
  CDFs]{\label{fig:aggregation_orders} Example of aggregation of $E=2$
    CDFs. The forecast CDFs (blue continuous line for expert $e=1$,
    red dashed line for expert $e=2$) are known only through a set of
    $M_1=M_2=3$ values $x_e^{m_e}$ with associated jumps
    $p_e^{m_e}=\frac{1}{3}$. Following Equation~\eqref{eq:aggregation_order}, since $x_2^1$
    is greater than $x_1^1$ and $x_1^2$, it is the quantile of order
    $\tau_{2;t}^1 =
    \omega_{1;t}(p_1^1+p_1^2)+\omega_{2;t}p_2^1=\frac{2}{3}\omega_{1;t}+\frac{1}{3}\omega_{2;t}$
    of $\widehat{y}_{t}$. If $\omega_{1;t}=\omega_{2;t}=\frac{1}{2}$,
    then $\tau_{2;t}^1=\frac{1}{2}$. In this case, the aggregated step-wise CDF is the black continuous line.}
\end{figure}

In other words, $x_{e;t}^{m_e}$ may be considered as the quantile of order
$\tau_{e;t}^{m_e}$ of some random variable $\widehat{Y}_t \in \mathcal{Y}$, whose computation is
illustrated in Figure~\ref{fig:aggregation_orders}. To produce a valid
step-wise CDF, the aggregation method must produce aggregation
weights that are all non negative and sum up to 1 at fixed $t$. Thus,
the aggregated forecast is a convex combination of the expert forecasts.

\subsection{Performance Assessment Tools}
\label{sec:aggr_verif}

Among the several attributes of performance of probabilistic forecast
systems, reliability and resolution are most useful
\citep{murphy1993good,winkler1996scoring}. \red{A probabilistic
  forecast system is reliable if the conditional distribution of the (random)
  observation $Y$ given the forecast distribution $F$, noted $[Y|F]$, is equal to the
  forecast distribution: $[Y|F] = F$.} \red{Resolution refers to the ability of the forecast system to issue forecast distributions very different from the marginal
distribution of the observations. If a forecast is reliable, decisions can
then be made by considering the observation will be drawn from the forecast distribution. For a reliable forecasting system, the higher is its resolution, the more useful it is useful for economical decision taking \citep{richardson2001measures}.} % In weather forecasting, if the observation distribution is
% stationary, always forecasting the climatological distribution
% (i.e. the marginal distribution of past observations) gives a
% perfectly reliable (but uninformative) forecast system. By definition,
% the climatological distribution, although reliable, has no resolution
% at all. A weather forecast system has a high resolution if it can
% issue forecast distributions very different from the climatological
% distribution by dynamically adapting its forecast to the
% meteorological situation at hand. 

\red{In order to compare the two approaches of probabilistic forecast selection, as
  stated in the Introduction, two tools are used: the Continuous
  Ranked Probability Score which is a global measure of performance, and
  the rank histogram which is linked to reliability.}

The Continuous Ranked Probability Score (CRPS,
\citeauthor{matheson1976scoring} \citeyear{matheson1976scoring}) is a
scoring rule that quantifies the forecast performance of a
probabilistic forecasts expressed as a CDF $F$ with a scalar
observation $y \in \mathcal{Y} \subseteq \mathbb{R}$
\begin{equation}
  \ell(F,y) =\int_{x\in\mathcal{Y}}\left(F(x)-H(x-y)\right)^2dx.
  \end{equation} The CRPS being convex in its first argument, the existence
  of theoretical bounds for the regret is ensured for some aggregation
  methods by the theory of prediction with expert advice. Since the experts are
  step-wise CDFs with steps at $x_{e;t}^{m_e}$, the information about the underlying forecast CDF $F$ is incomplete, which makes some estimators of the CRPS biased, as
  investigated in \cite{zamo2018estimation}. This last article shows that,
  in order to get an accurate estimate of the CRPS, one has to choose an estimator of the CRPS according to the nature of the $x_{e;t}^{m_e}$ (a sample from $F$ or a set of quantiles from $F$). This
  recommendation was followed in this study. \red{For a forecast CDF
    built as the empirical CDF of an $M$-random sample $x_{i=1,\ldots,M}$ (such as a raw
    ensemble), the CRPS is estimated with
\begin{equation}
\widehat{CRPS}_{INT}(M) = \frac{1}{M}\sum_{i=1}^M|x_i-y| - \frac{0.5}{M^2}\sum_{i,j=1}^M|x_i-x_j| \label{eq_crps_nrj_est}
\end{equation}
and for a forecast CDF defined from a set of $M$ quantiles
$x_{i=1,\ldots,M}$ with regularly spaced orders (such as a post-processed ensemble), the following estimator is used
\begin{align}
 \widehat{CRPS}_{PWM}(M) &= \frac{1}{M}\sum_{i=1}^M|x_i-y|  - \frac{0.5}{M(M-1)}\sum_{i,j=1}^M|x_i-x_j|  \label{eq_crps_pwm_est}
\end{align}}

\red{The expectation of
    the CRPS can be decomposed into three terms that quantify
    different properties of a forecasting system or the observation distribution: the
    reliability (RELI), the resolution (RES) and the  uncertainty (UNC) terms:
\begin{equation}
\mathbb{E}_{F,Y}\ell(F,Y) = RELI - RES + UNC
\end{equation}
Let us note $\bar{\pi} = [Y]$ the marginal distribution
  of the observation $Y$ and $\pi^F=[Y|F]$ the conditional distribution of the
  observation $Y$ given the forecast distribution $F$. Then the three terms are defined as
  in \cite{brocker2009reliability} by
\begin{align}
RELI & = & \mathbb{E}_{F,Y \sim \pi^F}\left(\ell(F, Y) - \ell(\pi^F, Y)\right) \\
RES & = & \mathbb{E}_{F,Y \sim \pi^F}\left(\ell(\bar{\pi}, Y) - \ell(\pi^F, Y)\right) \\
UNC & = & \mathbb{E}_{Y \sim \bar{\pi}}\ell(\bar{\pi}, Y)
\end{align}}
The reliability term is the average difference of CRPS between forecasts $F$ and $\pi^F$ when forecasting observations distributed according to $\pi^F$. It is negatively oriented with a minimum of 0 for a
perfectly reliable forecast system ($\pi^F=F\quad\forall F$). The resolution term is the average difference of CRPS between forecasts $\bar{\pi}$ and $\pi^F$ when forecasting observations distributed according to $\pi^F$. It is positively oriented and the more $F$ is different form $\bar\pi$ the higher it is. For a reliable forecast system, it is essentially equivalent to the sharpness. The uncertainty term is linked to the dispersion of the marginal distribution of the observation and does not depend on the forecast. \cite{hersbach2000decomposition} gives a method to estimate
these terms from an ensemble forecast. \red{Based on this
  decomposition, good CRPS can be obtained with a bad reliability
  (high RELI) provided that the resolution is good enough (high RES). By
  minimizing the CRPS as a forecast selection criterion, one may thus
  select a forecast system that has a low CRPS but is not reliable
  \citep{wilks2018enforcing}. The solution proposed by
  \cite{wilks2018enforcing} is to minimize the CRPS modified with a
  penalty proportional to the reliability term. Although this may
  indeed decrease the reliability term, there is no way to check if it is
  low enough.}

\red{\cite{collet2017generic} proposed a post-processing
  method that enforces the reliability of the forecast. To have a more
  general forecast selection method, it is proposed here to check that
  a forecasting system is reliable by testing the hypothesis that its
  rank histogram is flat. This procedure may be used as long as one
  can build a rank histogram from the forecast, which is generally the case.} The
rank histogram of an ensemble forecast, simultaneously introduced by
\cite{anderson1996method}, \cite{hamill1996random} and
\cite{talagrand1997evaluation}, is the histogram of the rank of the
observation when it is pooled with its corresponding forecast
members. For a reliable ensemble, the observation and the members must
have the same statistical properties, resulting in a flat rank
histogram. The type of deviation from flatness gives indications about the
flaws of an ensemble. For instance, an L-shape histogram means the
forecasts are consistently too high, while a J-shape histogram
indicates consistently too low forecasts. A U-shape histogram reveals
the forecast distribution is under-dispersed or conditionally
biased. \cite{hamill2001interpretation} showed on synthetic data that a
flat rank histogram can be obtained for an unreliable ensemble. Producing a
flat rank histogram is thus a necessary but not sufficient condition
for a forecasting system to be reliable. \red{Although the rank histogram was initially designed for ensemble forecasts, it can be used for forecasts in the form of quantiles. If the quantiles' orders are regularly spaced between 0 and 1 (excluded), then the histogram should also be flat if the forecasting system is reliable.} The flatness of a rank histogram
can be statistically tested thanks to the Jolliffe-Primo tests of
flatness described in \cite{jolliffe2008evaluating} and summarized in
Appendix~\ref{sec:JP_test}. These tests assess the existence of some
specific deviations from flatness in a rank histogram, such as a
slope, a convexity and a wave shape\footnote{The p-values of the Jolliffe-Primo test
  for slope and convexity have been computed based on a modified
  version of the function {\tt TestRankhist} in the R package {\tt
    SpecsVerification} \citep{SpecsVerification}. The function has
  been modified to compute also the p-value for the Jolliffe-Primo
  test for a wave shape, introduced in this
  study.}. \red{In this study we apply the three flatness tests to each rank histogram at several locations where forecasts and observations are available, which leads to a multiple testing procedure. In order to take into account this multiple testing, the false discovery rate is controlled thanks to the Benjamini-Hochberg procedure \citep{benjamini1995controlling,benjamini2001control} with an alpha of 0.01. \cite{brocker2018assessing} showed that when the observation ranks are serially dependent, the JP tests
  should be adapted to take into account this temporal dependency. In this
  study, the lag-1 autocorrelation of the rank has a median of 0.2
  over all the studied locations and lead times and is lower than 0.4
  for most of the forecasting systems (raw, post-processed and aggregated
  alike). This correlation was judged low enough so the modified
  procedure proposed by \cite{brocker2018assessing} was not used.}

\section{Aggregation Methods}
\label{sec:aggr_weights}

Five aggregation methods are introduced, from simple empirical ones
to more sophisticated ones derived from the theory of prediction with
expert advice. %For all but the sharpness-calibration aggregation method, the loss $\ell$ is based on the CRPS.

\subsection{Inverse CRPS Weighting}

  The \emph{inverse CRPS weighting method} (INV) gives to each
  expert's forecast a weight inversely proportional to its average
  CRPS over the last $W$ days:
  \begin{equation}
    \omega_{e;t} = \frac{(\overline{CRPS}_{e;t})^{-1}(W)}{\sum_{e=1}^{E}(\overline{CRPS}_{e;t})^{-1}(W)}
  \end{equation}
  where $\overline{CRPS}_{e;t}(W)$ is the average CRPS of expert $e$
  during the $W$ days before time $t$.

\subsection{Sharpness-Calibration Paradigm}

%% \cite{gneiting2007probabilistic} proposed that probabilistic forecast performance should be evaluated according to the paradigm of \emph{maximizing the sharpness of the predictive distributions subject to calibration}.
The \emph{sharpness-calibration paradigm} (SHARP) of \cite{gneiting2007probabilistic}
can be used as an aggregation method to select at each time one
expert. The aggregated forecast is the forecast of the expert whose
range of the central $90\%$ interval $IQ_{90}$, averaged over the last
$W$ days is the lowest, among the experts whose reliability term, as
computed in \cite{hersbach2000decomposition}, is lower than a chosen
threshold $Reli_{th}$ over the last $W$ days. The aggregated forecast
gives an aggregation weight of 1 to the sharpest reliable expert's forecast and
of 0 to the other experts' forecast:
\begin{equation}
  \omega_{e;t}=\mathbb{1}\left(e = \argmin_{\{e|Reli_{e;t}(W) <Reli_{th}\}}\overline{IQ_{90}}_{e;t}(W)\right),
\end{equation}
where $Reli_{e;t}(W)$ is the reliability term of expert $e$ over the
$W$ days before time $t$, and $\overline{IQ_{90}}_{e;t}(W)$ is the
average range of the interval between quantiles of orders 0.95 and
0.05, forecasted by the expert $e$ over the $W$ days before time
$t$. If no expert has a reliability term lower than the reliability
threshold $Reli_{th}$ over the last $W$ days, the aggregated forecast
is just the expert with the lowest mean CRPS over the last $W$ days.

The three following methods are derived from the theory of prediction
with expert advice, and bounds for the regret may be computed.

\subsection{Minimum CRPS}

The \emph{minimum CRPS method} (MIN) chooses the best recent expert in
terms of CRPS, that is, the aggregation weight is 1 for the expert
with the lowest average CRPS over the last $W$ days, and 0 for all the
other experts:
\begin{equation}
  \omega_{e;t}=\mathbb{1}(e=e_t^\star(W)),
\end{equation}
where $e_t^\star(W)$ is the index of the expert with the minimum
average CRPS during the last $W$ days. The reference class
$\mathcal{C}$ is the set of the $E$ available experts, so that the
oracle for this method is the expert with the lowest CRPS averaged
over the period $\{1,\ldots,T\}$, $e_T^\star(T)$. This aggregation
method is called ``follow-the-best-expert'' in
\cite{cesa2006prediction}, who prove that, under several assumptions
on the loss, the regret of the aggregated forecast relatively
to the oracle is $o(\ln(T))$.

\subsection{Exponential Weighting}

The \emph{exponentially weighted average forecaster} (EWA) computes the  aggregation weights as
  \begin{equation}
    \omega_{e;t} = \frac{exp\{-\eta CRPS_{e;t}(W)\}}{\sum_{e=1}^E exp\{-\eta {CRPS_{e;t}(W)\}}},
  \end{equation}
  where $\eta \in \mathbb{R}^+$ is called the learning rate and
  $CRPS_{e;t}(W)$ is the cumulative CRPS of expert $e$ over the last
  $W$ days. The highest the learning rate, the lowest the weight for a
  bad expert. \red{For very large learning rates $\eta$, EWA is practically
    equivalent to the MIN method.} The reference class
  $\mathcal{C}$ is the set of the $E$ available experts. If $W$ spans
  the whole period before $t$, that
  is, if $W =t-1$ days, EWA competes well with the best expert,
  in terms of average CRPS over the whole period $\{1,\ldots,T\}$,
  with a boudned regret relatively to this oracle \begin{equation}\label{eq:aggr_EWA}
    \sup R^\mathcal{C}_T \leq \frac{\ln E}{\eta} + \frac{\eta T}{8}B^2,
  \end{equation}
  where $B$ is the upper bound of the loss function (see proof in  Appendix~\ref{sec:aggr_app_demo_bound}). \red{The demonstration requires only the
    convexity of the loss $\ell$ in its first argument.} In practice, for
  an unbounded loss such as the CRPS, $B$ is the maximum of
  observations and expert forecasts over the whole period
  $t=1,\ldots,T$. Although the theoretical bounds exist only for
  $W=t-1$, one can use a shorter sliding window $W$ in order to make
  the aggregation weights change quickly over time. This may
  improve the forecast of an unstationary variable of
  interest.

\subsection{Exponentiated Gradient}

  The \emph{exponentiated gradient forecaster} (GRAD) weights the experts with
  \begin{equation}
    \omega_{e;t} = \frac{exp\{-\eta \partial_e CRPS_{t}(W)\}}{\sum_{e=1}^E exp\{-\eta \partial_e CRPS_{t}(W)\}},
  \end{equation}
  where $CRPS_{t}(W)$ is the cumulative CRPS of the GRAD aggregated forecast over the last $W$
  days, and
  $\partial_eCRPS_{t}(W)=\frac{\partial
    CRPS_{t}(W)}{\partial \omega_{e;t}}$. Using Appendix~\ref{sec:aggr_app_grad}  leads to

  \begin{align}
    \partial_eCRPS_{t}(W) = & \sum_{s=t-W}^{t-1}\frac{\partial CRPS}{\partial\ome}(\widehat{y}_s,y_s) \notag \\
    = &\sum_{s=t-W}^{t-1}\left\{\sum_{\me = 1}^{\Me} \pe |\xes - y_s| -\sum_{e^\prime = 1}^E \omeps\sum_{\mep=1}^{\Mep}\pep\xeps\right\}\notag \\
                          & -\sum_{s=t-W}^{t-1}\left\{\sum_{e^\prime = 1}^E \omeps\left(\sum_{\me=1}^{\Me}\sum_{\mep=1}^{\Mep}\pe \pep|\xes-\xeps|\right)\right\},
  \end{align}
  which generalizes Equation (5.13) of \cite{baudin2015prevision} to
  the case of the aggregation of step-wise CDFs with any number of
  steps. The rationale for using the gradient is that, if the gradient
  is positive for an expert over the past $W$ forecasts, increasing
  the weight would have increased the CRPS (and decreased the
  performance) of the aggregated forecast. So for incoming forecasts,
  giving this expert a low weight should improve the forecast
  performance. The reverse is true for a negative gradient.
  
  For this aggregation method, the reference class $\mathcal{C}$ is
  the set of convex combinations with constant weights over the whole
  period $\{1,\ldots,T\}$, that is, such that
$\widetilde\omega_{e;t} = \widetilde\omega_e,\forall e\in
\{1,\ldots,E\}$ and $t\in\{1,\ldots,T\}$. The oracle is the best, in
terms of cumulative CRPS, constant convex combination of experts. This
is usually a better oracle than the best expert. If $W = t-1$ days,
the following bound immediately follows from
\cite{mallet2007description} or \cite{baudin2015prevision}
  \begin{equation}
    \sup R^\mathcal{C}_T \leq \frac{\ln E}{\eta} + \frac{\eta T}{2}C^2,
  \end{equation}
  where $C=\textrm{max}_{t \in \{1,\ldots,T\},e\in\{1,\ldots,E\}}|\frac{\partial CRPS}{\partial\ome}(\widehat{y}_t,y_t)|$.
\red{Although the oracle for GRAD is better than for EWA, the bounds
  may be larger, so that EWA may actually perform better than GRAD. In
practice, one has to try both methods to know which one performs best.}

\section{The Experts and the Observation}
\label{sec:aggr_experts}

In this work, aggregation is applied to ensemble forecasts of the 10~m
wind speed over France. Forecasting surface wind is quite difficult
due to complex interactions between phenomena at different spatio-temporal
scales. For instance, wind speed is influenced by large scale
atmospheric structures such as cyclones and anticyclones, but also by
local orography and surface friction. Local atmospheric effects such
as downward drifts under convective clouds also contribute to the
direction and speed of wind. National weather services need to have
skillful wind speed forecasts to issue early warnings to the
population and civil security services. Also, wind speed influences
many economic activities, such as sailing, windpower generation,
construction,... that need good forecasts for decision making.

This section presents the 28 experts used in this study and the wind speed observation used to assess the forecasts. 

\subsection{The Four Experts based on TIGGE}
\label{subsec:TIGGE}
The International Grand Global Ensemble, formerly the THORPEX
Interactive Grand Global Ensemble (TIGGE), was an international
project aiming, among other things, to provide ensemble prediction
data from leading operational forecast centers
\citep{bougeault2010thorpex,swinbank2016tigge}. Although the TIGGE
data set\footnote{The TIGGE data set can be retrieved from the ECMWF at
  http://apps.ecmwf.int/datasets or from the Chinese Meteorological
  Administration at http://wisportal.cma.gov.cn/wis/} includes 10
ensemble NWP models, only the four ensemble models available on a
daily basis at M\'et\'eo-France have been retained (see
Table~\ref{tab:aggr_tigge}) as experts, so that the aggregation methods may later
be used in operations at Météo-France.

The TIGGE ensembles are available on a grid size is 0.5$^\circ$. Over France this amounts to a total of 267
grid points. The study period goes from the $1^{st}$ January, 2011 to the
$31^{st}$ December, 2014 (so $T=1461$ in the notations of
Section~\ref{sec:aggr_aggr}). The lead times go from 6~h to 54~h
depending on the ensemble, with a timestep of 6~h. The forecast are
done at 1800~UTC for lead times $h$ from 6~h to 48~h, with a time
step of 6~h. This implies that for experts based on CMC and ECMWF,
whose  runtime is 1200~UTC, the actual lead time is $h+6$ (see
Table~\ref{tab:aggr_tigge}).

Each ensemble is an expert whose forecast CDF $\widehat{F}_{e;t}$ is
the empirical CDF of the members associated with the same weight
$p_e^{m_e}=\frac{1}{M}$, where $M$ is the number of members in the
ensemble.

\begin{table}[!h]
  \caption{\label{tab:aggr_tigge}Ensembles from TIGGE used in this study, with some of their characteristics.}
  \centering
  {\footnotesize
  \begin{tabular}{p{0.4\columnwidth}|c|c|c}
    Weather service & Members & Hour of the run used (UTC) & Lead times\\
    \hline
    \hline
    Canadian Meteorological Center (CMC) & 21 & 1200 & 12h to 54h\\
    European Center for Medium-Range Weather Forecasts (ECMWF) & 51 & 1200 & 12h to 54h\\
    M\'et\'eo-France (MF) & 35 & 1800 & 6h to 48h\\
    US National Centers for Environmental Prediction (NCEP) & 21 & 1800 & 6h to 48h\\
  \end{tabular}}
\end{table}

\subsection{The Twenty Four Experts built with EMOS Methods}
\label{sec:aggr_calib}

Each ensemble is post-processed with two kinds of EMOS: non-homogeneous
regression (NR, \citeauthor{gneiting2005calibrated}
\citeyear{gneiting2005calibrated}, \citeauthor{hemri2014trends}
\citeyear{hemri2014trends}) and quantile random forest (QRF, \citeauthor{meinshausen2006quantile} \citeyear{meinshausen2006quantile}, \citeauthor{zamo2014benchmarkII} \citeyear{zamo2014benchmarkII}, \citeauthor{taillardat2016calibrated} \citeyear{taillardat2016calibrated}).

The forecast CDF produced with NR is parametric. Following
\cite{hemri2014trends}, the square root of the forecast wind speed
$\widehat{f}_t$ is supposed to follow a normal distribution truncated
at 0 
\begin{equation}
\sqrt{\widehat{f}_t} \sim \mathcal{N}^0(a+b\overline{x}_t,
c^2+d^2 sd_t)
\end{equation} where $\overline{x}_t$ and $sd_t$ are the mean and
standard deviation of the square-root of the associated ensemble
values, forecasted at time $t$. \red{The four paremeters ($a,
  b, c$ and $d$) are optimized by maximizing the
  log-likelihood\footnote{With the function {\tt optim} in R
    \citep{R}.} over the last $W_{tr}$ forecast days. In order to
  build several NR-post-processed ensembles with different reaction time
  to the raw ensemble's performance changes, five sizes of the training
  window $W_{tr}$ are used, as summarized in
  Table~\ref{tab:aggr_calib}. These parametric CDFs can not be
  used as such in the framework of step-wise CDF aggregation. A
  step-wise $\widehat{y}_{e;t}$ is built from the squared quantiles of orders
$\{0,\frac{1}{100},\ldots,\frac{99}{100},0.999\}$ of the parametric
CDF produced by NR. The last order is not $1$ in order to avoid
infinite values.}

QRF is a non parametric machine learning method that produces a set of quantiles of
chosen order. A random forest is a set of regression trees built on a
bootstrapped version of the training data \citep{zamo2016gridded}. While building each tree, the
observations are split in two most homogeneous groups (in terms of
variance of the observation), according to a splitting criterion over an explanatory variable. At each split (or
node) only a random subset of the complete set of explanatory
variables are tested, until a stopping criterion is reached (such as a
maximum number of final nodes, called leaves). When a forecast is
required, a step-wise CDF is produced by going down the forest with
the vector of explanatory variables, computing the step-wise CDF of
the observations associated to each leaf and averaging those CDFs. In
practice, one requests a set of quantile orders and gets the
corresponding quantiles\footnote{In the R
  packages {\tt quantregForest} or {\tt ranger}.}. \red{The
  requested quantile orders are the same as for NR, except the last
  order which is 1, because QRF cannot produce infinite values.}
Since the forecast CDF is a step-wise function, the obtained quantiles
may contain many ties, that are suppressed by linearily interpolating
between the points defining the CDF's steps, as explained in
\cite{zamo2018estimation}.

The post-processing and aggregation methods are trained separately for each of
the eight lead times and each grid point. \red{Since NR is a
  parametric method with only four parameters, it can be trained with
  few data, thus a sliding-window training period is used. QRF, being a
  non parametric method, requires more training data to define the
  many splits in each tree. It is thus trained with 4-fold
  cross-validation: each year is successively used as a test sample, the reminaing three being used as the training sample.}

\red{From each of the four ensembles
  seven experts are built: the raw ensemble, the QRF-post-processed ensemble
  and five NR-post-processed versions of the ensemble (each with a different size of the
  training window, Table~\ref{tab:aggr_calib}). Altogether, 28 experts are aggregated. }

\begin{table}[!h]
  \caption{\label{tab:aggr_calib}Summary description of the EMOS methods used to post-process each ensembe.}
  \centering
\begin{tabular}{p{0.3\columnwidth}|p{0.65\columnwidth}}
    \multicolumn{2}{c}{Quantile Regression Forest (QRF)} \\
    \hline
    \hline
    Forecast distribution & Non-parametric (set of quantiles). \\
    Explanatory variables & Control member, ensemble mean, ensemble 0.1 and 0.9 quantiles, month. \\
    Training method & 4-fold cross-validation (3 training years, 1 test year). \\
    Orders of the forecast quantiles & $0,\frac{1}{100},\ldots,\frac{99}{100},1.$\\
     \multicolumn{2}{c}{}\\
    \multicolumn{2}{c}{Non-homogeneous Regression (NR)} \\
    \hline
    \hline
    Forecast distribution &  Parametric (truncated normal distribution for the square-root of wind speed).\\
    Explanatory variables & Mean and standard deviation of the raw ensemble.\\
    Training method & Likelihood maximization on a sliding window over the $W_{tr}$ previous
    days. Five windows are used: $W_{tr}=7, 30, 90, 365, t-1$ days \\
    Orders of the forecast quantiles & $0,\frac{1}{100},\ldots,\frac{99}{100},0.999$.
\end{tabular}
\end{table}

\subsection{The Observation}

The observation is the 10 m average wind speed analysis built in
\cite{zamo2016gridded}, at 267 grid points over France. Those grid
points are the same as the ensemble datasets \red{presented in
  Section \ref{subsec:TIGGE}}.

\section{Results}
\label{sec:aggr_results}

The five aggregation methods presented in
Section~\ref{sec:aggr_weights} have been investigated, with different
values of the tuning parameters $W$ and $\eta$. The values
of the tuning parameters tested in the study are listed in
Table~\ref{tab:aggr_aggr_params}. When two parameters are listed, all
combinations have been tried.

\begin{table}[!h]
  \caption{\label{tab:aggr_aggr_params}Sets of tried values for the parameters of the aggregation methods. When two parameters are given, all the possible combinations of values have been tried.}
  \centering
\begin{tabular}{p{0.4\columnwidth}|p{0.5\columnwidth}}
    Aggregation method & Parameters' values \\
    \hline
    \hline
Minimum CRPS or Inverse CRPS & $W= 7, 15, 30, 90, 365, t-1$ days \\
    \hline
Sharpness-calibration & $W= 7, 15, 30, 90, 365, t-1$ days, $Reli_{th}=0.1$ m/s \\
    \hline
Exponentiated weighting or Exponentiated gradient weighting& $W=7, 15, 30, 90, 365, t-1$ days, $\eta=10^{-1.5}, 10^{-1},10^{-0.5}, 1, 10^{0.5}, 10^{1.5},10^{2}$\\
  \end{tabular}
\end{table}

Hereafter the best expert and the best aggregation method is selected
following the two approaches mentioned above: the minimization of the
average CRPS or the maximization of the proportion of grid points with
a rank histogram deemed flat by the three flatness tests. \red{The
  best forecasting system in  terms of minimum CRPS is called the most skillful. The  forecasting system which gets the maximum proportion of grid points with a flat rank histogram is called the most reliable.}

\subsection{CRPS and Reliability}

\begin{figure}[!ht]
  \centering
  \includegraphics[width=0.8\columnwidth]{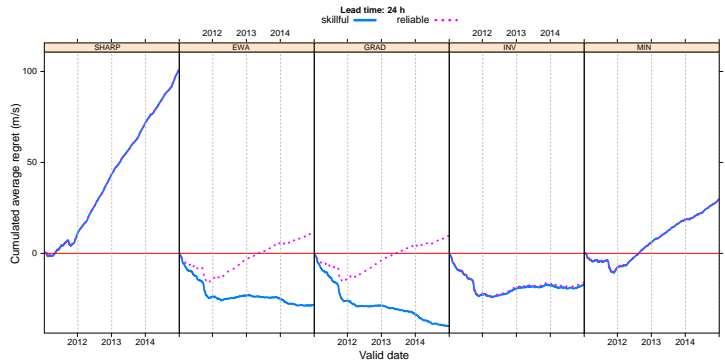}
  \caption[Time series of the cumulative average regret, at lead time
  24~h, for each aggregation
  method]{\label{fig:aggr_regret_ecmwf_qrf_h24}\red{Time series of the
    cumulative spatially-averaged regret, at lead time 24~h, for each aggregation
    method, relatively to the most skillful expert (QRF-post-processed
    ECMWF ensemble).  At each valid date, the regret relatively to the
    QRF-post-processed ECMWF ensemble is averaged over the 267
    grid-points. For each aggregation method, two settings are used to
    compute the regret: the most skillful one (blue continuous line)
    and the most reliable one (pink dashed line).}}
\end{figure}

From Table~\ref{tab:aggr_crps}, the most skillful expert is
the QRF-post-processed ECMWF ensemble (it is an oracle). The time series of the regret of
each most skillful or reliable aggregation method of each type
relatively to the QRF-post-processed ECMWF ensemble is drawn in
Figure~\ref{fig:aggr_regret_ecmwf_qrf_h24}, for lead time 24~h. Whereas
the most skillful setting of the SHARP aggregation method gets a consistently higher CRPS than the
most skillful expert, the other aggregation methods manage to
outperform the latter at least for some part of the four years. The
most skillful settings of EWA and GRAD even get a negative regret
relatively to the most skillful expert. The regret exhibits a trend
and a diurnal cycle with the lead times (not shown), and so does the
averaged CRPS (see Table~\ref{tab:aggr_crps}): the forecast performance decreases with
increasing lead times, particularly during the late afternoon when the
wind strengthens. The main point is that, in terms of CRPS, post-processing
improves performance, and aggregation further improves
performance. According to the minimization of the CRPS, the chosen
forecast method would be the most skillful GRAD setting, that is
$\log_{10}(\eta)=-1$ and $W=t-1$ days.

\begin{table}[ht]
\centering
\caption[Comparison of the CRPS averaged for the best expert and
aggregated forecasts]{\label{tab:aggr_crps}\red{CRPS averaged over the four
  years and the 267 grid points for several forecasting systems. The average CRPS of the most skillful
  raw ensemble (ECMWF) is indicated in the first line. Then, for each model selection approach, the best expert is indicated followed by the best setting of each aggregation method. For each selection approach,
  the forecasting system with the lowest CRPS is in bold.}}
\begin{tabular}{p{0.08\columnwidth}ccccccccccc}
%\rowcolor{Lightgray}
Method & \multicolumn{2}{c}{Parameters} & \multicolumn{9}{c}{Lead time (h)} \\
   \cline{2-3}
%   \rowcolor{Lightgray}
  & $\log_{10}(\eta)$ & $W$ & all & 6 & 12 & 18 & 24 & 30 & 36 & 42 & 48 \\ 
  \hline
  RAW ECMWF & & & \multirow{2}{*}{0.76} & \multirow{2}{*}{0.79} & \multirow{2}{*}{0.79} & \multirow{2}{*}{0.73} & \multirow{2}{*}{0.73} & \multirow{2}{*}{0.78} & \multirow{2}{*}{0.78} & \multirow{2}{*}{0.73} & \multirow{2}{*}{0.74} \\
                          % &&&&&&&&&&&\\
\rowcolor{Lightgray}
  \multicolumn{12}{c}{Selection: most skillful forecasting system.} \\
  \hline
  expert (QRF ECMWF) & & & \multirow{3}{*}{0.49} & \multirow{3}{*}{0.47} & \multirow{3}{*}{0.46} & \multirow{3}{*}{0.48} & \multirow{3}{*}{0.50} & \multirow{3}{*}{0.49} & \multirow{3}{*}{0.49} & \multirow{3}{*}{0.52} & \multirow{3}{*}{0.53} \\ 
SHARP &  & 1095 & 0.55 & 0.52 & 0.52 & 0.55 & 0.56 & 0.55 & 0.55 & 0.59 & 0.60 \\ 
\textbf{GRAD} & \textbf{-1} & \textbf{t-1} & \textbf{0.47} & \textbf{0.44} & \textbf{0.44} & \textbf{0.46} & \textbf{0.47} & \textbf{0.47} & \textbf{0.47} & \textbf{0.51} & \textbf{0.51} \\ 
EWA & -1 & 365 & 0.48 & 0.44 & 0.44 & 0.47 & 0.48 & 0.47 & 0.48 & 0.52 & 0.52 \\ 
INV & & 7 & 0.49 & 0.46 & 0.46 & 0.47 & 0.48 & 0.49 & 0.49 & 0.52 & 0.52 \\ 
MIN & &  365 & 0.51 & 0.47 & 0.47 & 0.51 & 0.52 & 0.50 & 0.51 & 0.56 & 0.56 \\ 
  % &&&&&&&&&&&\\
\rowcolor{Lightgray}
  \multicolumn{12}{c}{Selection: most reliable forecasting system.} \\
  \hline
  expert (QRF MF) & & & \multirow{3}{*}{0.50} &  \multirow{3}{*}{0.46} &  \multirow{3}{*}{0.47} &  \multirow{3}{*}{0.51} &  \multirow{3}{*}{0.50} &  \multirow{3}{*}{0.49} &  \multirow{3}{*}{0.50} &  \multirow{3}{*}{0.55} &  \multirow{3}{*}{0.53} \\
  SHARP & & 1095 & 0.55 & 0.52 & 0.52 & 0.55 & 0.56 & 0.55 & 0.55 & 0.59 & 0.60 \\ 
GRAD & 0.5 & t-1 & 0.50 & 0.46 & 0.46 & 0.50 & 0.50 & 0.49 & 0.50 & 0.54 & 0.54 \\ 
EWA & 0.5 & 30 & 0.50 & 0.46 & 0.46 & 0.50 & 0.50 & 0.50 & 0.50 & 0.55 & 0.54 \\ 
\textbf{INV} &  & \textbf{7} & \textbf{0.49} & \textbf{0.46} & \textbf{0.46} & \textbf{0.47} & \textbf{0.48} & \textbf{0.49} & \textbf{0.49} & \textbf{0.52} & \textbf{0.52} \\ 
MIN & & 365 & 0.51 & 0.47 & 0.47 & 0.51 & 0.52 & 0.50 & 0.51 & 0.56 & 0.56 \\ 
\end{tabular}
\end{table}

 % \red{But the choice of the most
 %  skillful aggregation method cannot be chosen from theoretical
 %  considerations. Although the theory of prediction with expert advice
 %  tells us that EWA performs more or less like a worst oracle than
 %  GRAD, it appears in this study that in practice EWA is almost as
 %  performant as GRAD. This may be different for other observation or
 %  geographical domain.}

\begin{figure}[!ht]
  \centering
  \subfloat[][]{
    \label{fig:aggr_rkhistmaps_raw}
    \includegraphics[width=0.47\columnwidth]{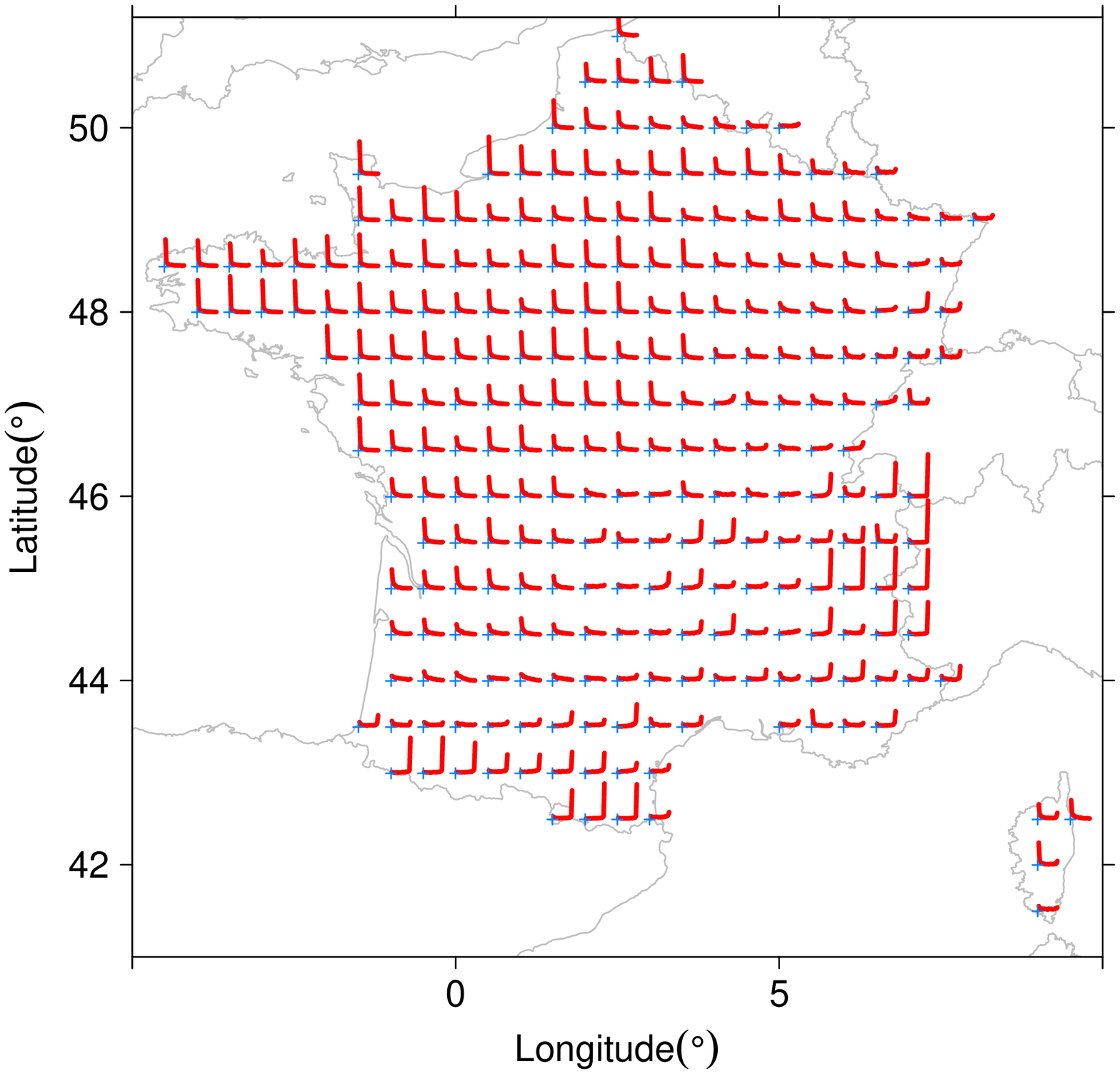}}
  \subfloat[][]{
    \label{fig:aggr_rkhistmaps_nrinf}
    \includegraphics[width=0.47\columnwidth]{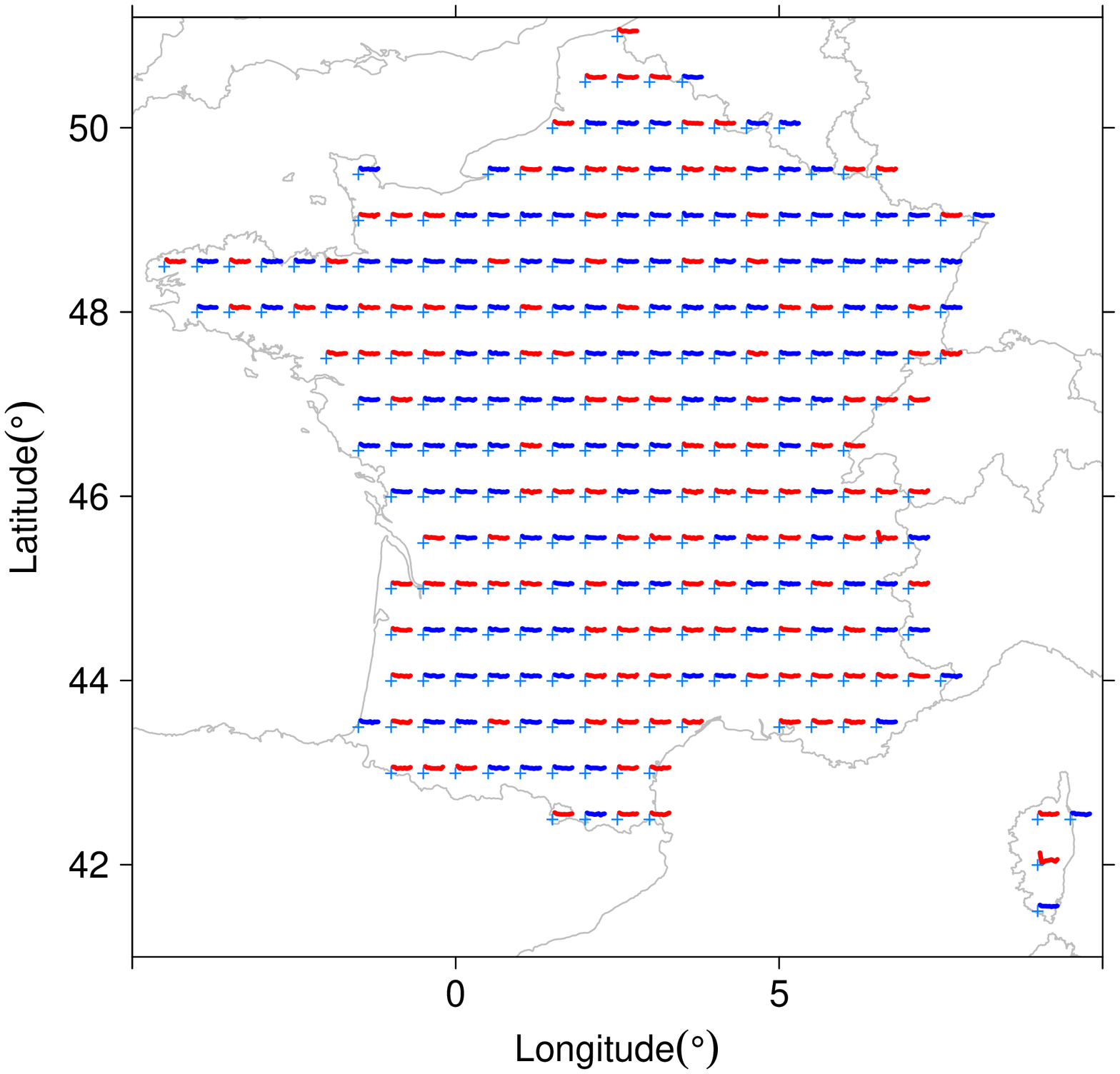}}\\
  \subfloat[][]{
    \label{fig:aggr_rkhistmaps_aggr}
    \includegraphics[width=0.47\columnwidth]{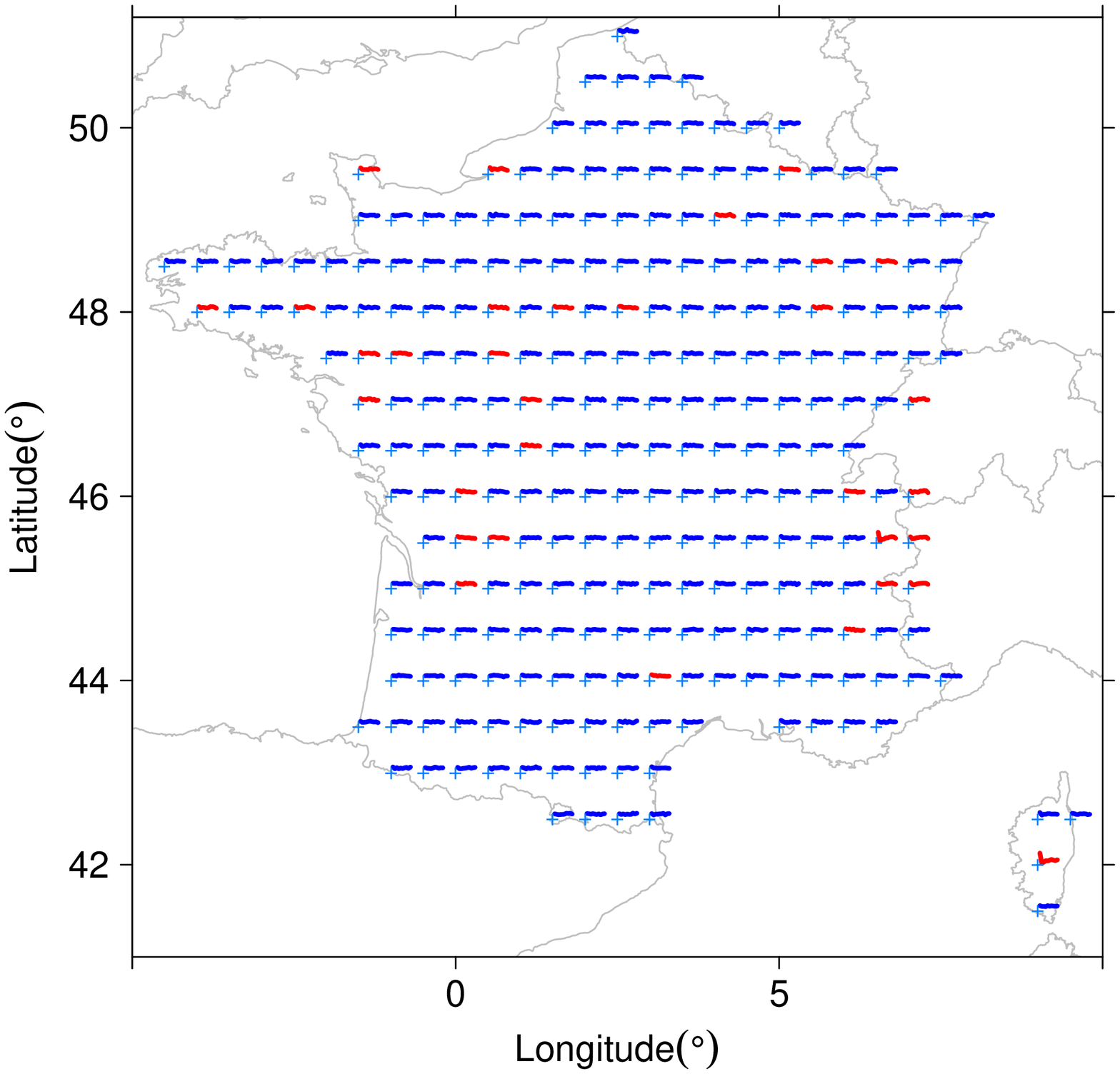}
  }
  \caption[Maps of sketches of the rank histograms]{\red{Maps of the rank
    histograms computed over the 4 years, for
    \subref{fig:aggr_rkhistmaps_raw} the raw CMC ensemble,
    \subref{fig:aggr_rkhistmaps_nrinf} the MF ensemble post-processed
    with QRF, and \subref{fig:aggr_rkhistmaps_aggr}
    EWA (with $\eta=10^{0.5}$ and $W=90$ days). Each grid-point is
    located by a blue cross. Rank histograms are represented as a line
    at the right of the associated grid-point, with the same vertical scale at
    all grid-points and for all maps. The lead time is $h=6$~h. A blue
    line means that none of the slope, convexity and wave JP tests
    rejects the flatness hypothesis, whereas a red line indicates at
    least one of the tests rejects the flatness hypothesis.}}
  \label{fig:aggr_rkhistmaps}
\end{figure}

If one uses the flatness of rank histograms as the selection
criterion, the best raw ensemble is CMC, the best expert is
QRF-post-processed MF and the best aggregated forecast is EWA with
$\log_{10}(\eta) = 0.5$ and $W=30$ days (see
Table~\ref{tab:aggr_jptests}). Raw CMC is not a very reliable
ensemble, as shown in
Figure~\ref{fig:aggr_rkhistmaps}~\subref{fig:aggr_rkhistmaps_raw} for
lead time $h=6$~h: the ensemble is consistently biased with too strong
forecast wind speeds in the north-west of France and too weak forecast
wind speeds over the Alps and the Pyrenees. Elsewhere, although
the ensemble is much less biased, the rank histogram is not deemed
flat due to an obvious U-shape. The rank histograms at the other lead
times and for the other raw ensembles exhibit similar features (not
shown). For the most reliable expert and aggregated forecasts, the
rank histograms are computed with the nine forecast deciles. As
illustrated in
Figure~\ref{fig:aggr_rkhistmaps}~\subref{fig:aggr_rkhistmaps_nrinf},
the QRF-post-processed version of the MF ensemble yields a higher number
of flat rank histograms than the raw CMC ensemble. Finally,
Figure~\ref{fig:aggr_rkhistmaps}~\subref{fig:aggr_rkhistmaps_aggr}
shows that, the JP tests do not reject the flatness
hypothesis at many more grid-points for the most reliable EWA
forecast. Table~\ref{tab:aggr_jptests} confirms quantitatively that
the most reliable EWA outperforms the most reliable expert in terms of
flatness of the rank histogram, at each lead time.% However a bad
                                % choice of the learning rate $\eta$
                                % and the aggregation window $W$ may
                                % actually decrease the number of flat
                                % histograms compared to the most
                                % reliable expert, as illustrated in
                                % Figure~\ref{fig:aggr_jptest_crps_exp}.
The most reliable setting of the other aggregation methods produce
fewer flat rank histograms than the most reliable expert, for all lead
times, as shown in Table~\ref{tab:aggr_jptests}. In other words,
post-processing improves reliability over raw ensemble, and aggregation
may further improve reliability over the most reliable expert if the
right setting is chosen.

\begin{table}[!t]
\centering
\caption[Comparison of the proportion of flat rank histograms for the
best expert or aggregated
forecasts.]{\label{tab:aggr_jptests}\red{Proportion of grid points
    where the three flatness tests do not reject the hypothesis of a
    flat rank histogram for several forecasting systems. For each model selection approach, the best expert is indicated followed by the best setting of each aggregation method. The forecasting system with the highest proportion of
    rank histograms is in bold, for each selection approach.}}
\begin{tabular}{L{0.08\columnwidth}ccccccccccc}
%\rowcolor{Lightgray}
  Method & \multicolumn{2}{c}{Parameters} & \multicolumn{9}{c}{Lead time (h)} \\
 \cline{2-3}
%\rowcolor{Lightgray}
         & $\log_{10}(\eta)$ & $W$ & all & 6 & 12 & 18 & 24 & 30 & 36 & 42 & 48 \\ 
  % \hline
  %   &&&&&&&&&&&\\
\rowcolor{Lightgray}
  \multicolumn{12}{c}{Selection: most skillful forecasting system.} \\
  \hline
expert (QRF ECMWF) & & & \multirow{3}{*}{0.60} & \multirow{3}{*}{0.26} & \multirow{3}{*}{0.17} & \multirow{3}{*}{0.83} & \multirow{3}{*}{0.89} & \multirow{3}{*}{0.41} & \multirow{3}{*}{0.43} & \multirow{3}{*}{0.95} & \multirow{3}{*}{0.86} \\
 SHARP & & 1095 & 0.39 & 0.28 & 0.22 & 0.39 & 0.36 & 0.53 & 0.45 & 0.45 & 0.41 \\ 
 GRAD & -1 & t-1 & 0.05 & 0.06 & 0.06 & 0.03 & 0.04 & 0.06 & 0.05 & 0.03 & 0.04 \\ 
 EWA & -1 & 365 & 0.04 & 0.04 & 0.03 & 0.03 & 0.04 & 0.06 & 0.06 & 0.05 & 0.04 \\ 
 INV & & 7 & 0.01 & 0.03 & 0.02 & 0.00 & 0.01 & 0.02 & 0.01 & 0.00 & 0.01 \\ 
 \textbf{MIN} & & \textbf{365} & \textbf{0.82} & \textbf{0.83} & \textbf{0.79} & \textbf{0.84} & \textbf{0.69} & \textbf{0.88} & \textbf{0.87} & \textbf{0.92} & \textbf{0.75} \\ 
  % &&&&&&&&&&&\\
\rowcolor{Lightgray}
  \multicolumn{12}{c}{Selection: most reliable forecasting system.} \\
  \hline
expert (QRF MF) & & &  \multirow{3}{*}{0.92} &  \multirow{3}{*}{0.76} &  \multirow{3}{*}{0.89} &  \multirow{3}{*}{0.99} &  \multirow{3}{*}{0.99} &  \multirow{3}{*}{0.87} &  \multirow{3}{*}{0.87} &  \multirow{3}{*}{1.00} &  \multirow{3}{*}{0.99} \\
  SHARP & & 1095 & 0.39 & 0.28 & 0.22 & 0.39 & 0.36 & 0.53 & 0.45 & 0.45 & 0.41 \\ 
 GRAD & 0.5 & t-1 & 0.67 & 0.46 & 0.43 & 0.91 & 0.79 & 0.46 & 0.51 & 0.98 & 0.79 \\ 
 \textbf{EWA} & \textbf{0.5} & \textbf{30} & \textbf{0.97} & \textbf{0.96} & \textbf{0.90} & \textbf{0.98} & \textbf{0.99} & \textbf{0.97} & \textbf{0.97} & \textbf{0.98} & \textbf{0.98} \\ 
 INV & & 7 & 0.01 & 0.03 & 0.02 & 0.00 & 0.01 & 0.02 & 0.01 & 0.00 & 0.01 \\ 
 MIN & & 365 & 0.82 & 0.83 & 0.79 & 0.84 & 0.69 & 0.88 & 0.87 & 0.92 & 0.75 \\ 
\end{tabular}
\end{table}

\red{The two selection approaches choose a different ``best''
  forecast. The GRAD aggregation method with $\log_{10}(\eta) = -1$
  and $W=t-1$ days is the most skillful forecast, whereas EWA with
  $\log_{10}(\eta) = 0.5$ and $W=30$ days is the most reliable
  forecast. In terms of averaged CRPS over all lead times, the most
  skillfull forecast is about 6~\% more performant than the most
  reliable forecast, as computed from Table~\ref{tab:aggr_crps}
  (0.47~m/s versus 0.50~m/s). Table~\ref{tab:aggr_jptests} shows that
  the most skillful forecast passes the three flatness tests for only
  $5\%$ of grid points, whereas the most reliable forecast produces about
  $97\%$ of flat rank histograms. In other words, selecting the most
  reliable forecast as best forecast, instead of the most skillful,
  leads to a slightly worse CRPS but increases dramatically the number of
  grid points with a flat rank histogram. This discrepancy between the
  two criteria for choosing the best forecasts is discussed more
  deeply in Section~\ref{sec:aggr_discussion}. For decision making, it
  is important and easier to have a reliable forecast, since this
  allows us to take the forecast as the true distribution of the observation and make optimal decision
  in terms of economic returns. Therefore, in the following the
  best retained forecast is EWA with $\log_{10}(\eta) = 0.5$ and
  $W=30$ days.}

\subsection{Temporal Variation of the Weights}

\begin{figure}[!ht]
  \centering
 \includegraphics[height=\columnwidth, angle=-90]{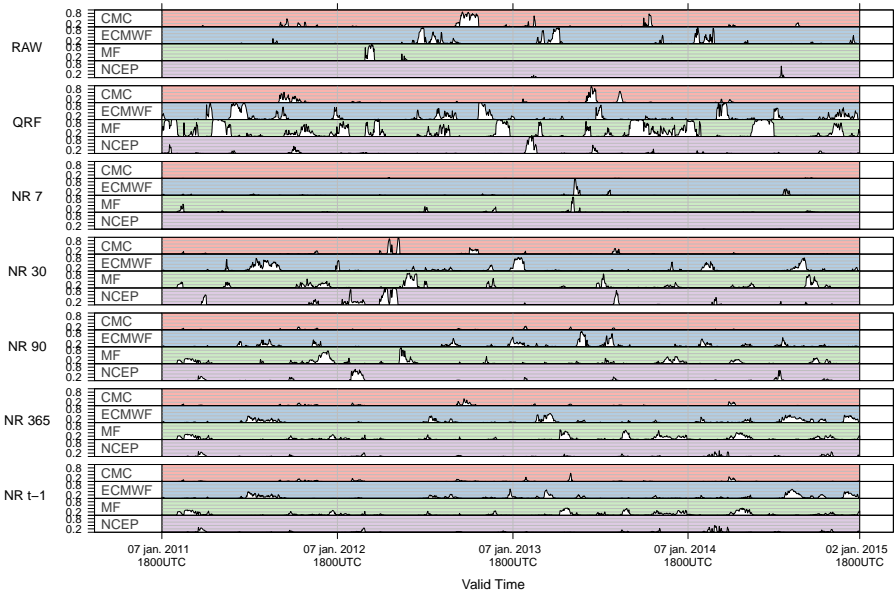}
\caption[Evolution of the aggregation weights with the valid date, for lead time 42~h]{\label{fig:weights}Evolution of the aggregation weights with the valid date, for lead time 42~h. The aggregation method is the EWA forecaster with $\log_{10}(\eta)=0.5$ and $W=30$ days.}
\end{figure}

The most reliable EWA aggregation is able to quickly redistribute the
aggregation weights between the experts, as illustrated in
Figure~\ref{fig:weights}. For instance, during the middle of 2011,
nearly all the aggregation weight shifts from the QRF-post-processed MF
ensemble to the QRF-post-processed ECMWF ensemble, in a few days. The
aggregation weights can also remain stable for long periods of time,
such as around mid-2014, when the QRF-post-processed MF ensemble keeps a
high weight for about two months. Moreover, although the raw ensembles
do not perform well on their own, the EWA method may find periods
where the raw ensembles can significantly contribute to the aggregated
forecasts, such as in late 2012 for the raw CMC ensemble. Last, the
time series of the aggregation weights is very different from one lead
time to another (not shown). These features prove that this
aggregation method is very adaptive. This may be very useful for
operations when an ensemble undergoes important changes: the
aggregation method will quickly detect a modification in performances
and adjust its weighting accordingly. 

% \subsection{Stability of the Choice of the Aggregation}

% The best aggregation methods and its parameters have been chosen
% \emph{a posteriori} over a period of four years, and not with an
% on-line choice \citep{gerchinovitz2008further,devaine2009aggregation}. In order to
% assess the variability of this choice, the proportion of flat rank
% histograms of the experts and the aggregated forecasts have been
% computed for each year separately. Whereas the most reliable expert
% varies from year to year, the most reliable aggregation is always EWA
% with $\eta=10^{0.5}$ and $W=30$ days, which is also the best set of
% values over the four years. Although the proportion of flat rank histograms is
% higher when computed over one year (about $90\%$) than over four years
% (about $80\%$), it gives a stable ranking of the aggregation
% methods. This allows to have a fair confidence in the future
% performance of the aggregated forecast, even after fitting its
% parameters over only one year.

\red{\subsection{Aggregation of individual sorted experts}}

\red{\cite{baudin2015prevision} and \cite{thorey2017thesis} defines
  experts as the sorted values of the pooled raw ensembles,
  and aggregate step-wise CDFs with one step. To compare our approach
  to their's, we aggregated the sorted values of the raw
  ensembles. This amounts to 128 experts. We also aggregated the
  sorted values of the raw and post-processed ensembles. This amounts to
  2,552 experts (the 128 values forecasted by the raw ensembes plus
  101 quantiles for each of the 24 post-processed forecasts). A
  comprehensive comparison with the approach in
  \cite{baudin2015prevision} being out of the scope of this article,
  only the most reliable aggregation method (EWA, with $\log_{10}(\eta)=0.5$
  and $W=30$ days) has been used.} 

\red{The resulting CRPSs are shown in Table \ref{tab:aggr_crps_1}. The forecast skill is improved compared to the raw ensembles (compare with
Table~\ref{tab:aggr_crps}). Even if no post-processed ensemble is used,
the CRPS is improved by the aggregation. But adding more experts
(among which are post-processed ensembles) further increases the performance
improvement. However, the CRPS stays higher than the CRPS of the EWA
aggregation of multi-step CDFs (0.52~m/s versus 0.50~m/s with our most
reliable EWA). Although the differences in CRPS is low, the difference between
the two approaches proves much more important in terms of proportion
of flat rank histograms. Whereas our most reliable EWA gets about 97\%
of flat rank histograms (see Table~\ref{tab:aggr_jptests}), the
aggregation with EWA of one-step CDFs gets no flat rank histogram
whatever lead time is considered. All the p-values of the
Jolliffe-Primo tests are lower than $2.10^{-25}$ (with a lag-1
autocorrelation of the rank around 0.12). Further investigations would
be required to check if this holds for other settings of EWA and for
other aggregation methods. A possible explanation of this lack of
reliability would be the fact that, for 1-member ensembles, the CRPS
reduces to the mean absolute error (MAE). Since the MAE is minimized by the
conditional median of the observation given the forecast,  each expert is weighted
according to its ability to predict well the conditional median,
which is not what is looked for.} 

\begin{table}[ht]
\centering
\caption[Comparison of the CRPS averaged for the best expert and
aggregated forecasts]{\label{tab:aggr_crps_1}\red{CRPS of the aggregated
  forecast built from experts considered as one-step CDFs. The
  experts are either the sorted members of the raw ensembles, or the
  sorted values of the raw and post-processed ensembles. The aggregation
  method is EWA with $\log_{10}(\eta) = 0.5$ and $W=30$ days.}}
\begin{tabular}{L{0.25\columnwidth}ccccccccc} 
% \rowcolor{Lightgray}
  \multirow{2}{*}{Aggregated experts} & \multicolumn{9}{c}{Lead time (h)} \\
  \cline{2-10}
% \rowcolor{Lightgray}
         & all & 6 & 12 & 18 & 24 & 30 & 36 & 42 & 48 \\ 
  % \hline
  % \multicolumn{12}{c}{Most reliable settings.} \\
%  \hline
Sorted raw members ($E = 128$)& \multirow{2}{*}{0.58} & \multirow{2}{*}{0.57} & \multirow{2}{*}{0.57} & \multirow{2}{*}{0.57} & \multirow{2}{*}{0.57} & \multirow{2}{*}{0.58} & \multirow{2}{*}{0.58} & \multirow{2}{*}{0.61} & \multirow{2}{*}{0.60} \\
\hline
Sorted raw and post-processed members ($E = 2,552$) & \multirow{4}{*}{0.52} & \multirow{4}{*}{0.48} & \multirow{4}{*}{0.49} & \multirow{4}{*}{0.50} & \multirow{4}{*}{0.51} & \multirow{4}{*}{0.52} & \multirow{4}{*}{0.53} &
\multirow{4}{*}{0.56} & \multirow{4}{*}{0.56} \\ 
\end{tabular}
\end{table}

\section{Discussion about Probabilistic Forecast Selection}
\label{sec:aggr_discussion}

The discrepancy between the choice of the best forecast according to
the CRPS and according to the flatness of rank histograms is now more fully
investigated and discussed.

\begin{figure}[!ht]
  \centering
  \includegraphics[width=0.68\columnwidth]{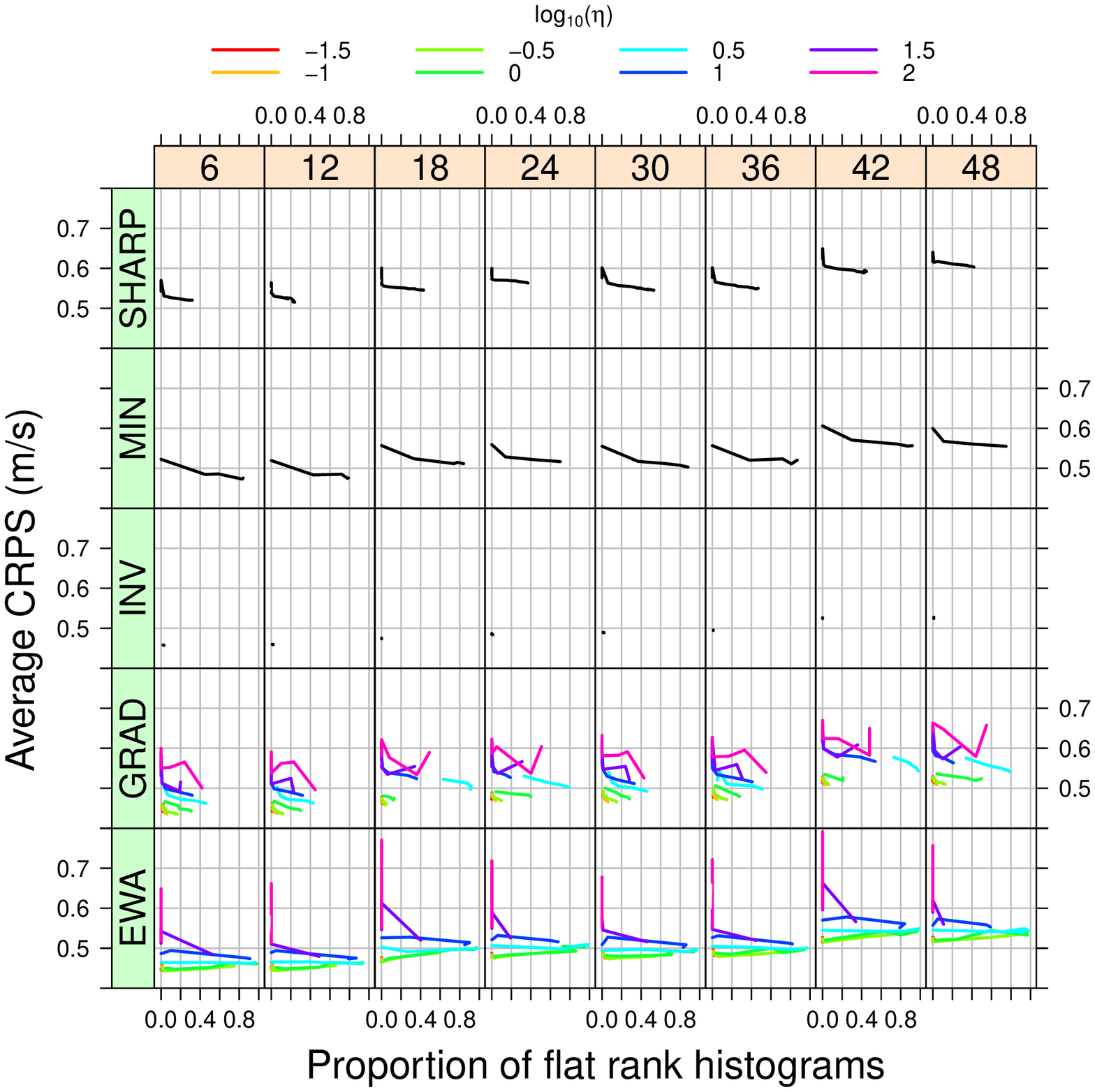}\\
  \includegraphics[width=0.68\columnwidth]{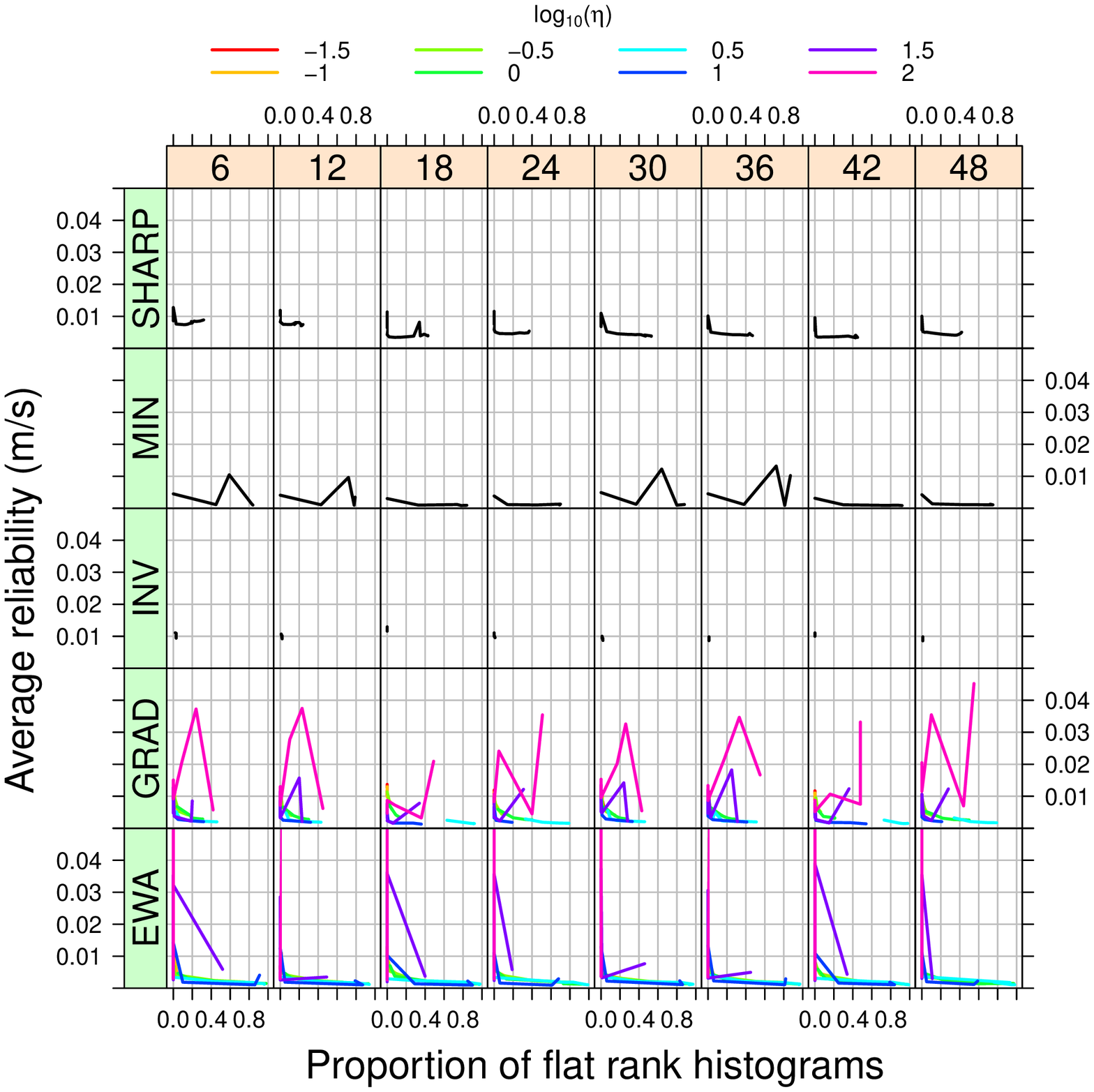}
  \caption[CRPS averaged versus the proportion of flat histograms, for
  the aggregations]{\label{fig:aggr_crpsvsjptest_aggrs}CRPS averaged
    over space and time (left) and reliability term (right), versus the
    proportion of flat rank histograms. \red{Each row corresponds to one
    aggregation method. Each column corresponds to one lead
    time. Inside each individual panel, the line is obtained by
    varying the aggregation window $W$. For EWA and GRAD, several
    lines are drawn, each associated with a different value of $\eta$
    (as indicated in the upper color legend).For INV, the variations are so small that it appears almost as a point.}} \end{figure}

Although the CRPS is a natural measure of performance for
forecast CDF, minimizing it does not ensure to get the highest number
of grid-points with a flat histogram, as stated earlier and as
confirmed in Figure~\ref{fig:aggr_crpsvsjptest_aggrs} (left). In this
figure, for each lead time, EWA and GRAD reach a minimum average CRPS
for very low proportions of flat rank histograms. Actually, the
point corresponding to this optimal CRPS is
indicated by the graph of INV in
Figure~\ref{fig:aggr_crpsvsjptest_aggrs} (left). \red{The graph for
  INV seems to reduce to a point because the associated CRPS and
  proportion of flat rank histograms barely vary.} For post-processed
ensembles, the same behavior can be observed (see Appendix~\ref{sec:CRPS_reli}): the most skillful expert may not be the most reliable
one. %. For
                                %instance, at lead time 36~h, the
                                %QRF-post-processed ECMWF ensemble
                                %minimizes the CRPS of the experts,
                                %but exhibits less than $40\%$ of flat
                                %histograms despite the low
                                %significance threshold chosen for the
                                %JP-test that allows important
                                %departure from flatness. At the same
                                %lead time, the NR-post-processed CMC
                                %ensemble (with $W=90$ days) has
                                %nearly $80\%$ of flat histograms for
                                %a slightly higher CRPS.

This discrepancy can be explained with the decomposition of the average CRPS as
a sum of a reliability term (that must be minimized) minus a
resolution term (that must be maximized) and an uncertainty term (that does not depend on the forecasting system). The reliability and the
resolution terms in the average CRPS after post-processing or aggregation
have very different ranges of variation from one setting to another,
as \red{deduced from} Figure~\ref{fig:aggr_crpsvsjptest_aggrs} for the aggregated forecasts. For
instance, at a fixed lead time, the reliability term of the MIN methods varies by less than
0.01 m/s while the CRPS varies by 0.05 m/s. Since the uncertainty term depends
only on the observation and is fixed for a fixed lead-time, the CRPS varies 
mainly because of the variations in the resolution term. In other words,
in this case, the average CRPS is mostly a measure of resolution and
its minimization as a selection criterion leads to maximize
the resolution. But the distributions may not be reliable since a
small change in the reliability term will not change a lot the average
CRPS while being compatible with unreliable flat rank histograms
according to the JP-tests (see right side of
Figure~\ref{fig:aggr_crpsvsjptest_aggrs}). This comment is also true
for the other aggregation methods. It also holds for post-processed ensembles, whose reliability term may change by about 0.01 m/s whereas the average CRPS
varies by 0.2 m/s with very different proportions of reliable grid
points (see Appendix~\ref{sec:CRPS_reli}).

\red{Another part of the origin of the discrepancy between the two
selection approaches is that the CRPS quantifies the
forecast performance over the whole distribution, while the
JP tests assess only forecast performance for the tested
shapes of the rank histogram. Firstly, the rank histogram takes only into account the
ranking of the observation and the members (or quantiles) while the relability term
of the CRPS takes into account the distance between the observation
and the members (or quantiles). Secondly, limiting the flatness criterion to the
absence of slope, convexity and wave shapes may lead to miss other
important deviations to flatness in the rank histogram. One could use
more flatness tests to add constraints on the forecasting system selected on a reliability criterion. It would be interesting investigate how it changes the solution.}

\red{Finally, the sampling noise may add further differences between
  the solutions selected by the two approaches. The CRPS and the
  flatness tests may not react in the same way to this sampling. The
  JP-test being hypothesis testing it has the same limit as every
  tests. Rejection of the flatness hypothesis may be due
  to other features than a lack of reliability \citep{ams2019inference}.}

In conclusion, the selection of the best forecast should not be made only
by minimizing the CRPS, but also by taking care of the actual
reliability of the forecasts, as tested with the JP
tests. Both performance critera should be used. Being only
focused on the CRPS may lead to choosing a forecasting system that is not
optimal in terms of reliabilty as shown here and earlier
studies \citep{collet2017generic,wilks2018enforcing}. But relying only
on the JP flatness tests may also be misleading. Indeed,
always forecasting the climatology leads to a very good reliability
but a very low resolution. Consequently, a forecaster may be tempted
to forecast the climatology in order to get a good forecast, instead
of issuing a forecast he might consider more likely but too different
from the climatology, and too risky to issue. Choosing a forecast
based on the JP tests only may lead to such hedging strategies. In
this study, hedging is avoided by using the CRPS to tune the experts
and the aggregation. But the reliability of the chosen forecast is
ensured by using the JP tests.

% Instead of choosing the most skillful post-processing or aggregation
% method, retaining the most reliable one comes with a very moderate
% increase in CRPS due to a slightly lower resolution term (compare the
% most skillful and most reliable EWA in Table~\ref{tab:aggr_crps}),
% while greatly increasing the number of flat rank histograms (again,
% compare the most skillful and most reliable EWA in
% Table~\ref{tab:aggr_jptests}). 

\section{Conclusion and Perspectives}
\label{sec:aggr_conclusion}

\red{The first goal of the present study was to adapt the theory of
prediction with expert advice to the case of experts issuing
probabilistic forecasts as step-wise CDFs with any number of
steps.} Contrary to the work of \cite{baudin2015prevision} who
aggregated unidentifiable experts built by pooling and sorting members
of several ensembles, each expert used in the present work is identifiable over time as required by the
theoretical framework of prediction with expert advice: the aggregation weights for the members or quantiles of the
same expert are constrained to be equal.  Some formulae of \cite{baudin2015prevision}
valid for step-wise CDFs with one step have been generalized to the
case of step-wise CDFs with any number of steps. 

Several aggregation methods to combine step-wise forecast CDFs have
been presented and compared in terms of reliability and CRPS. The
reliability has been assessed by using the Jolliffe-Primo tests, which
detect the presence in the rank histogram of typical deviations from
flatness. The systematic use of the Jolliffe-Primo flatness test
highlights that the minimization of the CRPS as the main criterion to
calibrate or aggregate may not produce the maximum number of flat rank
histograms. It is also shown that choosing the best forecast by
maximizing the proportion of rank histograms ensures reliable
forecasts, without significantly increasing the CRPS.

On a real wind speed data set, the best aggregation method, in terms
of proportion of flat rank histograms, is the exponentially weighted
average forecaster, with a learning rate $\eta=10^{0.5}$ and an
aggregation window $W=30$ days. This aggregated forecast has a similar
CRPS as the most skillful expert in terms of CRPS, and produces many
more flat rank histograms than the most reliable expert. The method
can produce weights with very different temporal patterns: rapidly
evolving weighting of the experts, long period of constant weighting,
short period with large weights for the raw ensembles. With the use of
experts fitted over sliding windows of different size, this
flexibility may help to solve a recurrent problem in post-processing:
important changes in the NWP models that may make the post-processing
equation inadequate for the new version of the NWP model. \red{Although EWA
  is the selected aggregation method in this study, this must not be
  taken as a result valid for other observations and/or geographical
  domains. To the best of our knowledge, no theoretical results allow us
  to tell among the many available aggregation methods which one will
  be the best on a given dataset.}

As for the perspectives, it is planned to study the same and other
aggregation methods by pooling data in blocks of nearby
grid-points. This may improve the fit by enlarging the training sample
or, at the very least, speed up operations on finer grids with
thousands of points, as was demonstrated for deterministic
forecasts \citep{zamo2016gridded}. Since post-processing of other
meteorological parameters, such as temperature and rainfall, has been
already tested internally at M\'et\'eo-France, aggregation methods
will be tried on these variables too. A more comprehensive study of
the discrepancy between the CRPS and the proportion of flat rank
histograms as a performance criterion and its implication on
post-processing and aggregation constitutes a more theoretical
perspective. A further line of research would be to expand on the
proposed approach to post-processing and aggregation. Indeed, the retained
criterion in this study (maximizing the number of grid points with a
flat rank histogram) does not allow us to choose a different forecast for
each grid point, which may be desirable to further improve forecast
performance.

\FloatBarrier

\appendix
\section*{Appendix}

\section{Statistical Tests of Flatness of a Rank Histogram}
\label{sec:JP_test}

Consider the vector $\boldsymbol{\delta}$ of normalized deviation from flatness in each rank
of some rank histogram at hand,
  $$\boldsymbol{\delta}=\left(\frac{n_1-n_0}{\sqrt{n_0}},\ldots,\frac{n_k-n_0}{\sqrt{n_0}}\right),$$
  where $k$ is the number of possible ranks, $n_i$ is the count of
  rank $i$ and $n_0=\frac{\sum_{i=1}^k n_i}{k}$ is the theoretical
  count in each rank for a flat histogram.

Under the null hypothesis $H_0$ that the rank histogram is compatible with a
flat histogram up to sampling noise, the squared norm
$||\boldsymbol{\delta}||^2$ follows a $\chi^2$ distribution with $k-1$
degrees of liberty. The flatness of the rank histogram can be tested
with a chi-square test.

\begin{figure}[!t]
  \centering
  \includegraphics[width=0.6\columnwidth]{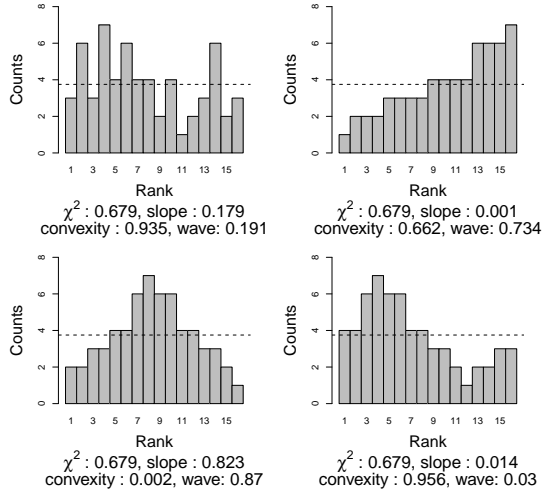}
  \caption[Illustration of the Jolliffe-Primo flatness
  tests]{\label{fig:aggr_chi2}Illustration of the Jolliffe-Primo
    flatness tests of a rank histogram. The p-value of four flatness
    tests (chi-square, and three Jolliffe-Primo tests) are reproduced.}
\end{figure}

 The chi-square test statistic $||\boldsymbol{\delta}||^2$ is
insensitive to the shape of the deviations to a flat histogram, as
shown in Figure~\ref{fig:aggr_chi2}. To build this figure, as in
\cite{elmore2005alternatives}, 60 integer values from 1 to 16 have
been drawn from a uniform distribution. Four histograms are shown: the
histogram computed with the raw sample (top left), with the same
counts sorted in ascending order (top right), with the counts
reassigned to have a peak-shaped histogram (bottom left), and with the
counts reassigned in a wave shape (bottom right). The p-value of the
chi-square test and three other flatness tests presented below is
reproduced under the histograms. Although the counts of each rank are
reorganized, the p-value of the chi-square test of the four histograms
is the same, because reordering the counts is equivalent to reordering
the components of $\boldsymbol\delta$, which does not change its norm. Because of
this, in our study, the flatness of each
rank histogram is assessed with the decomposition of the chi-square
test statistic, as detailed in \cite{jolliffe2008evaluating}. Under
the null hypothesis $H_0$, any
projection of $\boldsymbol{\delta}$ onto an orthonormal basis of
$\mathbb{R}^k$ has $k-1$ components whose squares are asymptotically
independent $\chi^2$ random variables, each with 1 degree of
freedom. If the basis vectors are chosen to describe a sloped
histogram, a convex histogram, or any other shape of interest, the
existence of the shape in the rank histogram can be tested. The
existence of a shape is not rejected if the projection of
$\boldsymbol{\delta}$ onto the corresponding basis vector has a
component statistically different from 0.

\cite{jolliffe2008evaluating} give formulae to compute the basis
vectors for deviations from flatness commonly encountered on real
data. As an example, if $k=2p+1$, the basis vector for the slope
(resp. convexity) test is proportional to $(-p,-p+1,\ldots,-p+(k-1))$
(resp. $(p^2-p\frac{(p+1)}{3},(p-1)^2-p\frac{(p+1)}{3},\ldots,-p\frac{(p+1)}{3},1-p\frac{(p+1)}{3},\ldots,p^2-p\frac{(p+1)}{3})$). In
our study, three tests are used, the slope and convexity tests, and
the ``wave'' test not described in \cite{jolliffe2008evaluating}. This
last test assesses the presence of a deviation from flatness in the
shape of a tilde, that was frequently observed in the literature
\citep{scheuerer2015probabilistic,baran2016mixture,taillardat2016calibrated}
and in internal studies at M\'et\'eo-France. The corresponding basis
vector is built thanks to the Grahm-Schmidt process as follows: the
vector $(0, \textrm{sin}(\frac{2\pi}{k-1}),
\textrm{sin}(2\pi\frac{2}{k-1}),\ldots,\textrm{sin}(2\pi\frac{k-2}{k-1}),0)$
is made orthogonal to the slope basis vector, and the resulting vector
is normalized to get the basis vector for testing the presence of a
wave shape. In Figure~\ref{fig:aggr_chi2}, the p-values for the test
of existence of a slope, a convexity or a wave are in agreement with
the shape of the histograms. For instance, the low p-value of the slope
test (top right) rejects flatness against slope as expected.

\section{Proof of the Bounds for the Regret of the Exponentially Weighted Average Forecaster}
\label{sec:aggr_app_demo_bound}

The proof closely follows the proof of theorem 2.2 in \cite{cesa2006prediction}.

Let $\ell:\widehat{\mathcal{Y}}\times\mathcal{Y}\rightarrow[a;b]$ be a real-valued, bounded loss function. $\ell$ is supposed convex in its first argument.

The EWA weights at time $t$ are computed as
\begin{equation*}
    \omega^{EWA}_{e;t} = \frac{exp\{-\eta L_{e;t}\}}{\sum_{e=1}^E exp\{-\eta L_{e;t}\}},
  \end{equation*}
  with $L_{e;t}=\sum_{s=1}^{t-1}\ell(\widehat{y}_{e;s},y_s)$ the cumulative loss of expert $e$ at time $t$, and with the convention $L_{e;1}=0$ so that $\omega^{EWA}_{e,1}=\frac{1}{E}\quad\forall e$.

  Let us define $W_t=\sum_{e=1}^Eexp\{-\eta L_{e;t}\} \forall t \ge 1$ and $W_0=E$. At all times $t = 1, \ldots, T$, and using the convention that a sum over 0 elements is 0 (for $t=1$, such that $\omega^{EWA}_{e,0}=\frac{1}{E}\quad\forall e$),
  \begin{align}
    \ln\frac{W_t}{W_{t-1}} = & \ln\frac{\sum_{e=1}^Eexp\{-\eta \ell(\widehat{y}_{e;t},y_t)\}exp\{-\eta L_{e;t-1}\}}{\sum_{e^\prime=1}^Eexp\{-\eta L_{e^\prime;t-1}\}}  \notag \\
    = & \ln\frac{\sum_{e=1}^E\omega^{EWA}_{e;t-1}exp\{-\eta \ell(\widehat{y}_{e;t},y_t)\}}{\sum_{e^\prime=1}^E\omega^{EWA}_{e^\prime;t-1}}. \label{eq:aggr_wt}
  \end{align}

  The proof now needs Hoeffding's inequality \citep{hoeffding1963probability}. Let $a,b \in \mathbb{R}$ with $a<b$. Let $Z$ be a bounded random variable with values in $[a;b]$, then, $\forall s \in \mathbb{R}$, Hoeffding's inequality states that
  \begin{equation*}
    \ln \mathbb{E}\left[e^{sZ}\right] \leq s\mathbb{E}[Z] + \frac{s^2}{8}(b-a)^2.
  \end{equation*}
  Using Equation~\eqref{eq:aggr_wt} and Hoeffding's inequality for the random variable $Z$ taking the values $\ell(\widehat{y}_{e;t},y_t)$ with discrete probability $\omega^{EWA}_{e;t-1}$, taking $s=-\eta$ and summing over $t=1,\ldots,T$ leads to
  \begin{align*}
    \ln\frac{W_T}{W_0} \le & -\eta\sum_{t=1}^T\sum_{e=1}^E\omega^{EWA}_{e;t}\ell(\widehat{y}_{e;t},y_t) + \frac{\eta^2}{8}(b-a)^2T \\
    \le & -\eta\sum_{t=1}^T\ell\left(\sum_{e=1}^E\omega^{EWA}_{e;t}\widehat{y}_{e;t},y_t\right) + \frac{\eta^2}{8}(b-a)^2T \\
                           = & -\eta\sum_{t=1}^T\ell\left(\widehat{y}_t,y_t\right) + \frac{\eta^2}{8}(b-a)^2T,
  \end{align*}
  after using the convexity of the loss function $\ell$ in its first argument, and the definition of the EWA forecast.

  Noting that the following relationship also holds
  \begin{align*}
    \ln\frac{W_T}{W_0} = & \ln\left(\sum_{e=1}^Eexp\{-\eta L_{e;T}\}\right) - \ln E\\
    \ge & \ln\left(\underset{e=1,\ldots,E}{\max} exp\{-\eta L_{e;T}\}\right) - \ln E\\
    = & -\eta \underset{e=1,\ldots,E}{\min} L_{e;T} - \ln E,
  \end{align*}
  and combining it with the previous relationship leads to
  \begin{align*}
-\eta \underset{e=1,\ldots,E}{\min} L_{e;T} - \ln E \le -\eta\sum_{t=1}^T\ell\left(\widehat{y}_t,y_t\right) + \frac{\eta^2}{8}(b-a)^2T.
  \end{align*}
 Finally, dividing by $-\eta$ results in the following bound of the regret of the aggregated forecast relatively to the best expert
  \begin{align}
\sum_{t=1}^T\ell\left(\widehat{y}_t,y_t\right) - \underset{e=1,\ldots,E}{\min} L_{e;T} \le \frac{\ln E}{\eta} + \frac{\eta}{8}(b-a)^2T.
  \end{align}

Noting that this bound for the regret holds for any bounded loss
function $\ell$ convex in its first argument, which is a propery of
the CRPS, concludes the demonstration.

\section{Formula for the Gradient of the CRPS}
\label{sec:aggr_app_grad}

\cite{baudin2015prevision} considers the aggregation of step-wise CDFs with one single step ($\Me = 1 \quad \forall e \in \{1,\ldots,E\}$). We generalize equations (5.10) and (5.13) of \cite{baudin2015prevision} for, respectively, the CRPS and gradient thereof, to an aggregation of step-wise CDFs with any number of steps.

Dropping the time index $t$ in the notations, the aggregated CDF at time $t$ is
\begin{align}
\widehat{y}(x) & = \sum_{e=1}^E\ome\left[\sum_{m_e=1}^{M_e}\pe H_{\xe}(x)\right],\notag
\end{align}
with the notation $H_a(x)=H(x-a)$.

Therefore, the CRPS of the aggregated CDF at time $t$ is
\begin{align}
  CRPS(\widehat{y},y) & = \int_{\mathbb{R}}\left\{ H_y(x) - \sum_{e=1}^E\ome\left[\sum_{m_e=1}^{M_e}\pe H_{\xe}(x)\right]\right\}^2dx \notag \\
  & = \int_{\gamma}^\Gamma\left\{ H_y(x) - \sum_{e=1}^E\ome\left[\sum_{m_e=1}^{M_e}\pe H_{\xe}(x)\right]\right\}^2dx, \notag
\end{align}
where $\gamma = \textrm{min} (y,x_{1}^{1},\ldots,x_{E}^{M_E})$ and $\Gamma = \textrm{max} (y,x_{1}^{1},\ldots,x_{E}^{M_E})$.

By developing the square inside the integral,
\begin{align}
  CRPS(\widehat{y},y) = & \int_{\gamma}^\Gamma H_y(x)dx \notag \\
                        & - 2 \int_{\gamma}^\Gamma\sum_{e=1}^E\ome\left[\sum_{m_e=1}^{M_e}\pe H_{\xe}(x)H_y(x)\right]dx \notag \\
                        & + \int_{\gamma}^\Gamma\left\{\sum_{e=1}^E\ome\left[\sum_{m_e=1}^{M_e}\pe H_{\xe}(x)\right]\right\}\left\{\sum_{e^\prime=1}^{E^\prime}\omep\left[\sum_{\mep =1}^{M_{e^\prime}}\pe H_{\xep}(x)\right]\right\}dx. \notag
\end{align}

Noting that $H_a(x)H_b(x) = H_{\textrm{max}(a,b)}(x)$, and $\int_\gamma^\Gamma H_a(x) dx = \Gamma- a \quad \forall a \in [\gamma;\Gamma]$, then
\begin{align}
  CRPS(\widehat{y},y) = & \Gamma - y \notag \\
                        & -2 \sum_{e=1}^E\ome\left\{\sum_{m_e=1}^{M_e}\pe [\Gamma-\textrm{max}(\xe, y)] \right\} \notag \\
                        & + \sum_{e,e^\prime = 1}^E \ome \omep \left\{\sum_{m_e=1}^{M_e}\sum_{m_{e^\prime}=1}^{M_{e^\prime}}\pe \pep \left[\Gamma - \textrm{max}(\xe, \xep)\right]\right\} \notag \\
  = & - y \notag \\
                        & + 2 \sum_{e=1}^E\ome\left[\sum_{m_e=1}^{M_e}\pe \textrm{max}(\xe, y) \right] \notag \\
                        & - \sum_{e,e^\prime = 1}^E \ome \omep \left\{\sum_{m_e=1}^{M_e}\sum_{m_{e^\prime}=1}^{M_{e^\prime}}\pe \pep \textrm{max}(\xe, \xep)\right\} \label{eq:aggr_crps},\notag
\end{align}
because $\sum_{e=1}^E\ome =1$, and $\sum_{m_e=1}^{M_e} \pe = 1 \quad \forall e \in \{1,\ldots,E\}$.

Since $\textrm{max}(a,b) = \frac{1}{2}\left(a+b+|a-b|\right)$,
\begin{align}
  CRPS(\widehat{y},y) = & - y \notag \\
                        & + \sum_{e=1}^E\ome\left[\sum_{\me = 1}^{\Me} \pe (\xe + y +|\xe - y|) \right]\notag \\
                        & -\frac{1}{2} \sum_{e, e^\prime = 1}^E \ome\omep\left\{\sum_{\me=1}^{\Me}\sum_{\mep=1}^{\Mep}\pe \pep(\xe+\xep+|\xe-\xep|)\right\}\notag \\
  = &\sum_{e=1}^E\ome\left[\sum_{\me = 1}^{\Me} \pe (\xe +|\xe - y|) \right]\notag \\
                        & -\frac{1}{2} \sum_{e, e^\prime = 1}^E \ome\omep\left\{\sum_{\me=1}^{\Me}\sum_{\mep=1}^{\Mep}\pe \pep(\xe+\xep+|\xe-\xep|)\right\}.
\end{align}

The derivation with respect to $\ome$ results in
\begin{align}
  \frac{\partial CRPS}{\partial\ome}(\widehat{y},y) = & \sum_{\me = 1}^{\Me} \pe (\xe +|\xe - y|)\notag \\
                        & -\sum_{e^\prime = 1}^E \omep\left\{\sum_{\me=1}^{\Me}\sum_{\mep=1}^{\Mep}\pe \pep(\xe+\xep+|\xe-\xep|)\right\}.\notag
\end{align}

Finally, recalling that $\sum_{e = 1}^E \ome = 1$, and $\sum_{\me = 1}^{\Me} \pe = 1 \quad \forall e \in \{1,\ldots,E\}$
\begin{align}
  \frac{\partial CRPS}{\partial\ome}(\widehat{y},y) = & \sum_{\me = 1}^{\Me} \pe |\xe - y| -\sum_{e^\prime = 1}^E \omep\sum_{\mep=1}^{\Mep}\pep\xep\notag \\
                        & -\sum_{e^\prime = 1}^E \omep\left\{\sum_{\me=1}^{\Me}\sum_{\mep=1}^{\Mep}\pe \pep|\xe-\xep|\right\}.\label{eq:aggr_crps_grad}
\end{align}

Formulae \eqref{eq:aggr_crps} and \eqref{eq:aggr_crps_grad} generalize equations (5.10) and (5.13), respectively, of \cite{baudin2015prevision}.

\section{Relationship between the experts' CRPS and the reliability term with
  the JP tests}
\label{sec:CRPS_reli}

\begin{figure}
  \centering
  \includegraphics[width=0.47\columnwidth]{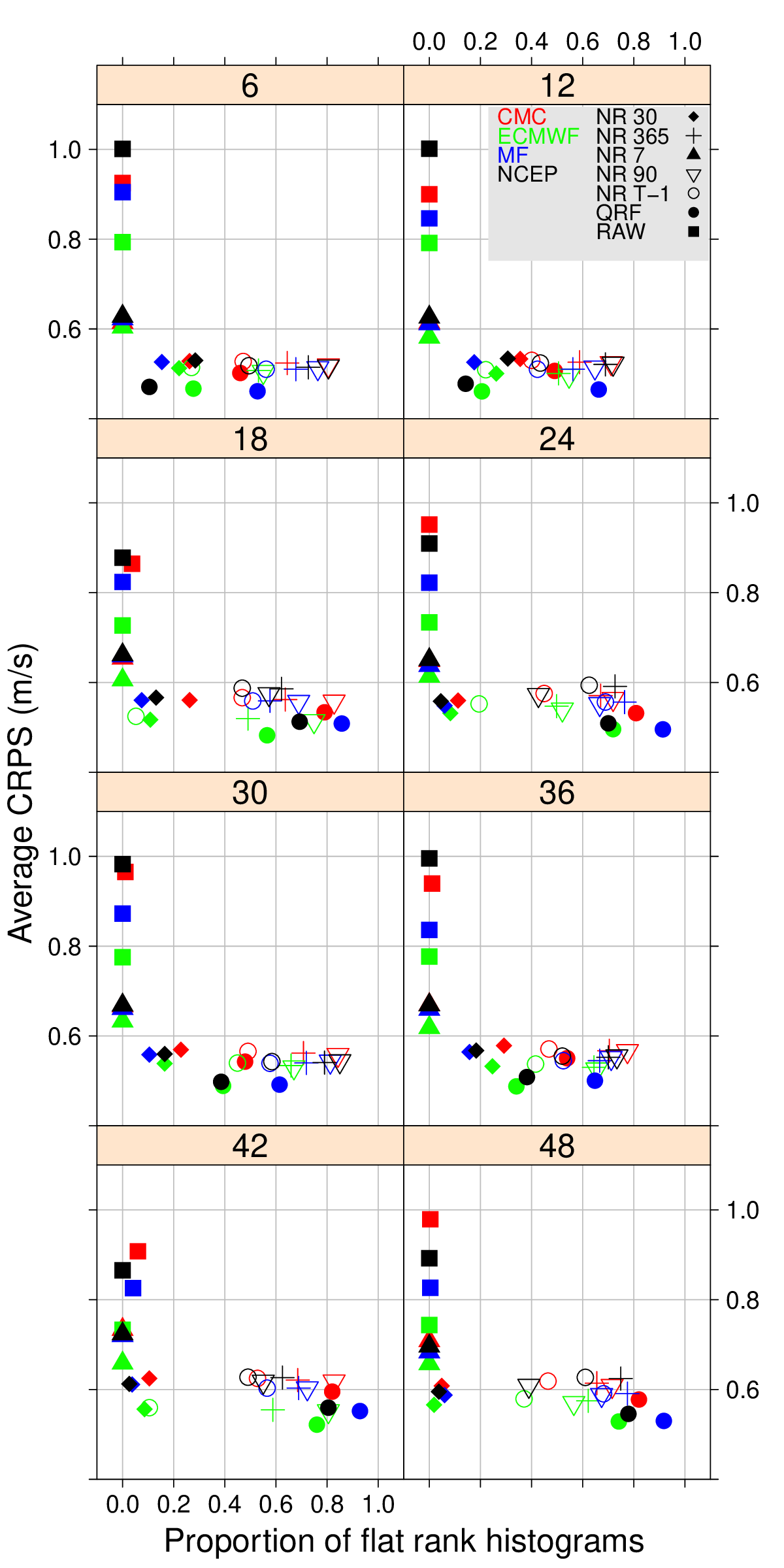}
%\caption[CRPS averaged versus the proportion of flat histograms, for the experts]{\label{fig:aggr_crpsvsjptest_experts}Same as
%  Fig.~\ref{fig:aggr_crpsvsjptest_aggrs} for the experts.}
%\end{figure}
%\begin{figure}
%  \centering
  \includegraphics[width=0.47\columnwidth]{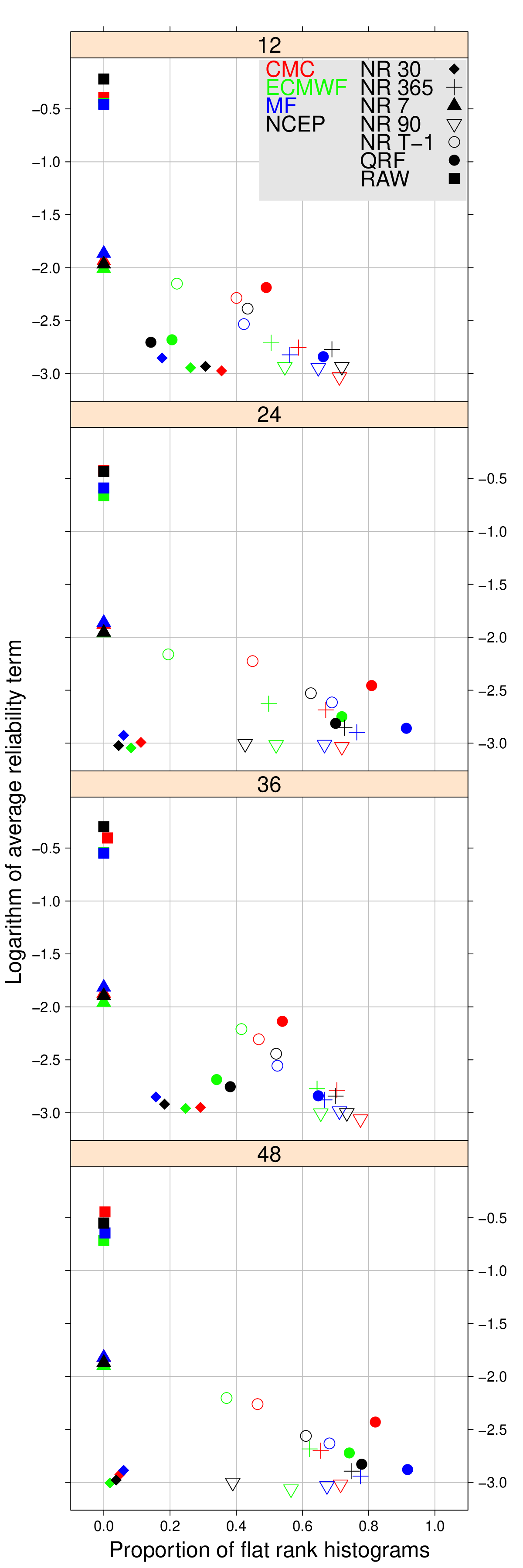}
\caption{\label{fig:aggr_crpsvsjptest_experts}CRPS averaged over space and time (left) and reliability term
  (righ), versus the proportion of flat rank histograms, for each
  expert, by lead time.}
\end{figure}

The expert with the minimum CRPS may not be the most reliable,
according to the proportion of flat rank histograms in view of the JP
tests. For instance in Figure~\ref{fig:aggr_crpsvsjptest_experts}, at
lead time 36~h, the QRF-post-processed ECMWF ensemble has the lowest CRPS
among the experts, but exhibits less than $40\%$ of flat histograms
despite the low significance threshold chosen for the JP tests that
allows important departure from flatness. At the same lead time, the
NR-post-processed version of the CMC ensemble (with $W=90$ days) has nearly $80\%$ of
flat histograms for a slightly higher CRPS.

As for the aggregation methods, the CRPS of the post-processed ensembles is
mainly driven by the resolution term. Whereas the reliability term
varies by less than 0.01~m/s, the average CRPS may change by up to
0.1~m/s from one to the other. For post-processed ensembles, the
average CRPS is thus mostly a measure of resolution only.

\FloatBarrier

% Submissions are not required to reflect the precise reference formatting of the journal (use of italics, bold etc.), however it is important that all key elements of each reference are included.
\bibliography{references}

\begin{thebibliography}{}

\bibitem[Adjakossa et~al., 2020]{adjakossa2020kalman}
Adjakossa, E., Goude, Y., and Wintenberger, O. (2020).
\newblock {Kalman Recursions Aggregated Online}.
\newblock {\em arXiv preprint arXiv:2002.12173}.

\bibitem[Allard et~al., 2012]{allard2012probability}
Allard, D., Comunian, A., and Renard, P. (2012).
\newblock Probability aggregation methods in geoscience.
\newblock {\em Mathematical Geosciences}, 44(5):545--581.

\bibitem[Anderson, 1996]{anderson1996method}
Anderson, J.~L. (1996).
\newblock A method for producing and evaluating probabilistic forecasts from
  ensemble model integrations.
\newblock {\em Journal of Climate}, 9(7):1518--1530.

\bibitem[Baran and Lerch, 2015]{baran2015log}
Baran, S. and Lerch, S. (2015).
\newblock {Log-normal distribution based Ensemble Model Output Statistics
  models for probabilistic wind-speed forecasting}.
\newblock {\em Quarterly Journal of the Royal Meteorological Society}.

\bibitem[Baran and Lerch, 2016]{baran2016mixture}
Baran, S. and Lerch, S. (2016).
\newblock {Mixture EMOS model for calibrating ensemble forecasts of wind
  speed}.
\newblock {\em Environmetrics}, 27:116--130.

\bibitem[Baudin, 2015]{baudin2015prevision}
Baudin, P. (2015).
\newblock {\em Pr{\'e}vision s{\'e}quentielle par agr{\'e}gation d'ensemble:
  application {\`a} des pr{\'e}visions m{\'e}t{\'e}orologiques assorties
  d'incertitudes}.
\newblock PhD thesis, Universit{\'e} Paris-Saclay.

\bibitem[Benjamini and Hochberg, 1995]{benjamini1995controlling}
Benjamini, Y. and Hochberg, Y. (1995).
\newblock Controlling the false discovery rate: a practical and powerful
  approach to multiple testing.
\newblock {\em Journal of the Royal statistical society: series B
  (Methodological)}, 57(1):289--300.

\bibitem[Benjamini and Yekutieli, 2001]{benjamini2001control}
Benjamini, Y. and Yekutieli, D. (2001).
\newblock The control of the false discovery rate in multiple testing under
  dependency.
\newblock {\em Annals of statistics}, 29(4):1165--1188.

\bibitem[Bogner et~al., 2017]{bogner2017combining}
Bogner, K., Liechti, K., and Zappa, M. (2017).
\newblock Combining quantile forecasts and predictive distributions of
  streamflows.
\newblock {\em Hydrology and Earth System Sciences}, 21(11):5493--5502.

\bibitem[Bougeault et~al., 2010]{bougeault2010thorpex}
Bougeault, P., Toth, Z., Bishop, C., Brown, B., Burridge, D., Chen, D.~H.,
  Ebert, B., Fuentes, M., Hamill, T.~M., Mylne, K., et~al. (2010).
\newblock {The THORPEX interactive grand global ensemble}.
\newblock {\em Bulletin of the American Meteorological Society}, 91(8):1059.

\bibitem[Br{\"o}cker, 2009]{brocker2009reliability}
Br{\"o}cker, J. (2009).
\newblock Reliability, sufficiency, and the decomposition of proper scores.
\newblock {\em Quarterly Journal of the Royal Meteorological Society},
  135(643):1512--1519.

\bibitem[Br{\"o}cker, 2018]{brocker2018assessing}
Br{\"o}cker, J. (2018).
\newblock Assessing the reliability of ensemble forecasting systems under
  serial dependence.
\newblock {\em Quarterly Journal of the Royal Meteorological Society},
  144(717):2666--2675.

\bibitem[Buizza et~al., 2005]{buizza2005comparison}
Buizza, R., Houtekamer, P., Pellerin, G., Toth, Z., Zhu, Y., and Wei, M.
  (2005).
\newblock {A comparison of the ECMWF, MSC, and NCEP global ensemble prediction
  systems}.
\newblock {\em Monthly Weather Review}, 133(5):1076--1097.

\bibitem[Cesa-Bianchi et~al., 2006]{cesa2006prediction}
Cesa-Bianchi, N., Lugosi, G., et~al. (2006).
\newblock {\em Prediction, learning, and games}, volume~1.
\newblock Cambridge University Press Cambridge.

\bibitem[Collet and Richard, 2017]{collet2017generic}
Collet, J. and Richard, M. (2017).
\newblock {A Generic Method for Density Forecasts Recalibration}.
\newblock In {\em Forecasting and Risk Management for Renewable Energy}, pages
  147--166. Springer.

\bibitem[Descamps et~al., 2011]{descamps2011representing}
Descamps, L., Labadie, C., and Bazile, E. (2011).
\newblock {Representing model uncertainty using the multiparametrization
  method.}
\newblock In {\em Proceedings of ECMWF Workshop on Representing Model
  Uncertainty and Error in Numerical Weather and Climate Prediction Models,
  20-24 June 2011}, pages 175--182.

\bibitem[Descamps et~al., 2014]{descamps2014pearp}
Descamps, L., Labadie, C., Joly, A., Bazile, E., Arbogast, P., and C{\'e}bron,
  P. (2014).
\newblock {PEARP, the M{\'e}t{\'e}o-France short-range ensemble prediction
  system}.
\newblock {\em Quarterly Journal of the Royal Meteorological Society},
  141(690):1671--1685.

\bibitem[Elmore, 2005]{elmore2005alternatives}
Elmore, K.~L. (2005).
\newblock Alternatives to the chi-square test for evaluating rank histograms
  from ensemble forecasts.
\newblock {\em Weather and forecasting}, 20(5):789--795.

\bibitem[Gerchinovitz et~al., 2008]{gerchinovitz2008further}
Gerchinovitz, S., Mallet, V., and Stoltz, G. (2008).
\newblock A further look at sequential aggregation rules for ozone ensemble
  forecasting.
\newblock {\em Rapport technique, INRIA Paris-Rocquencourt et {\'E}cole normale
  sup{\'e}rieure, Paris}.

\bibitem[Gneiting et~al., 2007]{gneiting2007probabilistic}
Gneiting, T., Balabdaoui, F., and Raftery, A. (2007).
\newblock Probabilistic forecasts, calibration and sharpness.
\newblock {\em Journal of the Royal Statistical Society: Series B (Statistical
  Methodology)}, 69(2):243--268.

\bibitem[Gneiting et~al., 2005]{gneiting2005calibrated}
Gneiting, T., Raftery, A., Westveld~III, A., and Goldman, T. (2005).
\newblock {Calibrated probabilistic forecasting using ensemble model output
  statistics and minimum CRPS estimation}.
\newblock {\em Monthly Weather Review}, 133(5):1098--1118.

\bibitem[Gneiting et~al., 2013]{gneiting2013combining}
Gneiting, T., Ranjan, R., et~al. (2013).
\newblock Combining predictive distributions.
\newblock {\em Electronic Journal of Statistics}, 7:1747--1782.

\bibitem[Hamill, 2001]{hamill2001interpretation}
Hamill, T. (2001).
\newblock Interpretation of rank histograms for verifying ensemble forecasts.
\newblock {\em Monthly Weather Review}, 129(3):550--560.

\bibitem[Hamill and Colucci, 1998]{hamill1998evaluation}
Hamill, T. and Colucci, S. (1998).
\newblock {Evaluation of Eta-RSM ensemble probabilistic precipitation
  forecasts}.
\newblock {\em Monthly Weather Review}, 126(3):711--724.

\bibitem[Hamill and Colucci, 1996]{hamill1996random}
Hamill, T.~M. and Colucci, S.~J. (1996).
\newblock {Random and systematic error in NMC’s short-range Eta ensembles}.
\newblock In {\em Preprints, 13th Conf. on Probability and Statistics in the
  Atmospheric Sciences, San Francisco, CA, Amer. Meteor. Soc}, pages 51--56.

\bibitem[Hemri et~al., 2014]{hemri2014trends}
Hemri, S., Scheuerer, M., Pappenberger, F., Bogner, K., and Haiden, T. (2014).
\newblock Trends in the predictive performance of raw ensemble weather
  forecasts.
\newblock {\em Geophysical Research Letters}, 41(24):9197--9205.

\bibitem[Hersbach, 2000]{hersbach2000decomposition}
Hersbach, H. (2000).
\newblock Decomposition of the continuous ranked probability score for ensemble
  prediction systems.
\newblock {\em Weather and Forecasting}, 15(5):559--570.

\bibitem[Hoeffding, 1963]{hoeffding1963probability}
Hoeffding, W. (1963).
\newblock Probability inequalities for sums of bounded random variables.
\newblock {\em Journal of the American statistical association},
  58(301):13--30.

\bibitem[Holton and Hakim, 2012]{holton2012introduction}
Holton, J.~R. and Hakim, G.~J. (2012).
\newblock {\em An introduction to dynamic meteorology}, volume~88.
\newblock Academic press.

\bibitem[Jolliffe and Stephenson, 2011]{jolliffe2011fverification}
Jolliffe, I. and Stephenson, D. (2011).
\newblock {\em Forecast verification: a practioner's guide in atmospheric
  science, second edition}.
\newblock Wiley-Blackwell.

\bibitem[Jolliffe and Primo, 2008]{jolliffe2008evaluating}
Jolliffe, I.~T. and Primo, C. (2008).
\newblock Evaluating rank histograms using decompositions of the chi-square
  test statistic.
\newblock {\em Monthly Weather Review}, 136(6):2133--2139.

\bibitem[Leutbecher and Palmer, 2008]{leutbecher2008ensemble}
Leutbecher, M. and Palmer, T.~N. (2008).
\newblock {Ensemble forecasting}.
\newblock {\em Journal of Computational Physics}, 227(7):3515--3539.

\bibitem[Malardel, 2005]{malardel2005fondamentaux}
Malardel, S. (2005).
\newblock {\em Fondamentaux de m{\'e}t{\'e}orologie: {\`a} l'{\'e}cole du
  temps}, volume~45.
\newblock Cepadues.

\bibitem[Mallet et~al., 2007]{mallet2007description}
Mallet, V., Mauricette, B., and Stoltz, G. (2007).
\newblock {Description of Sequential Aggregation Methods and their Performances
  for Ozone Ensemble Forecasting}.
\newblock {\em Technical report, \'Ecole normale sup\'erieure, DMA and CEREA}.

\bibitem[Matheson and Winkler, 1976]{matheson1976scoring}
Matheson, J.~E. and Winkler, R.~L. (1976).
\newblock Scoring rules for continuous probability distributions.
\newblock {\em Management science}, 22(10):1087--1096.

\bibitem[Meinshausen, 2006]{meinshausen2006quantile}
Meinshausen, N. (2006).
\newblock Quantile regression forests.
\newblock {\em The Journal of Machine Learning Research}, 7:983--999.

\bibitem[M{\"o}ller and Gro{\ss}, 2016]{moller2016probabilistic}
M{\"o}ller, A. and Gro{\ss}, J. (2016).
\newblock Probabilistic temperature forecasting based on an ensemble
  autoregressive modification.
\newblock {\em Quarterly Journal of the Royal Meteorological Society},
  142(696):1385--1394.

\bibitem[M{\"o}ller and Scheuerer, 2013]{moller2013postprocessing}
M{\"o}ller, D. and Scheuerer, M. (2013).
\newblock {\em {Postprocessing of Ensemble Forecasts for Wind Speed over
  Germany}}.
\newblock PhD thesis, Diploma thesis, Faculty of Mathematics and Computer
  Science, Heidelberg University. Available online at
  http://www.rzuser.uni-heidelberg.de/\~{}kd4/files/Moeller2013.pdf.

\bibitem[Murphy, 1993]{murphy1993good}
Murphy, A. (1993).
\newblock What is a good forecast? an essay on the nature of goodness in
  weather forecasting.
\newblock {\em Weather and Forecasting}, 8(2):281--293.

\bibitem[Mylne, 2002]{mylne2002decision}
Mylne, K.~R. (2002).
\newblock Decision-making from probability forecasts based on forecast value.
\newblock {\em Meteorological Applications}, 9(3):307--315.

\bibitem[{R Core Team}, 2015]{R}
{R Core Team} (2015).
\newblock {\em {R: A Language and Environment for Statistical Computing}}.
\newblock R Foundation for Statistical Computing, Vienna, Austria.

\bibitem[Richardson, 2001]{richardson2001measures}
Richardson, D.~S. (2001).
\newblock Measures of skill and value of ensemble prediction systems, their
  interrelationship and the effect of ensemble size.
\newblock {\em Quarterly Journal of the Royal Meteorological Society},
  127(577):2473--2489.

\bibitem[Scheuerer et~al., 2015]{scheuerer2015probabilistic}
Scheuerer, M., M{\"o}ller, D., et~al. (2015).
\newblock Probabilistic wind speed forecasting on a grid based on ensemble
  model output statistics.
\newblock {\em The Annals of Applied Statistics}, 9(3):1328--1349.

\bibitem[Siegert, 2015]{SpecsVerification}
Siegert, S. (2015).
\newblock {\em {SpecsVerification: Forecast Verification Routines for the SPECS
  FP7 Project}}.
\newblock R package version 0.4-1.

\bibitem[Stoltz, 2010]{stoltz2010agregation}
Stoltz, G. (2010).
\newblock Agr{\'e}gation s{\'e}quentielle de pr{\'e}dicteurs: m{\'e}thodologie
  g{\'e}n{\'e}rale et applications {\`a} la pr{\'e}vision de la qualit{\'e} de
  l'air et {\`a} celle de la consommation {\'e}lectrique.
\newblock {\em Journal de la Soci{\'e}t{\'e} Fran{\c{c}}aise de Statistique},
  151(2):66--106.

\bibitem[Swinbank et~al., 2016]{swinbank2016tigge}
Swinbank, R., Kyouda, M., Buchanan, P., Froude, L., Hamill, T.~M., Hewson,
  T.~D., Keller, J.~H., Matsueda, M., Methven, J., Pappenberger, F., et~al.
  (2016).
\newblock {The TIGGE project and its achievements}.
\newblock {\em Bulletin of the American Meteorological Society}, 97(1):49--67.

\bibitem[Taillardat et~al., 2016]{taillardat2016calibrated}
Taillardat, M., Mestre, O., Zamo, M., and Naveau, P. (2016).
\newblock {Calibrated Ensemble Forecasts using Quantile Regression Forests and
  Ensemble Model Output Statistics}.
\newblock {\em Monthly Weather Review}, 144(6):2375--2393.

\bibitem[Talagrand et~al., 1997]{talagrand1997evaluation}
Talagrand, O., Vautard, R., and Strauss, B. (1997).
\newblock Evaluation of probabilistic prediction systems.
\newblock In {\em Proc. ECMWF Workshop on Predictability}, pages 1--25.

\bibitem[Thorarinsdottir and Gneiting, 2010]{thorarinsdottir2010probabilistic}
Thorarinsdottir, T.~L. and Gneiting, T. (2010).
\newblock Probabilistic forecasts of wind speed: ensemble model output
  statistics by using heteroscedastic censored regression.
\newblock {\em Journal of the Royal Statistical Society: Series A (Statistics
  in Society)}, 173(2):371--388.

\bibitem[Thorey, 2017]{thorey2017thesis}
Thorey, J. (2017).
\newblock {\em {Pr\'evision d’ensemble par agr\'egation s\'equentielle
  appliqu\'ee \`a la pr\'evision de production d’\'energie
  photovolta\"ique.}}
\newblock PhD thesis, INRIA.
\newblock to be submitted (in English).

\bibitem[Wasserstein et~al., 2019]{ams2019inference}
Wasserstein, R., Schirm, A., and N., L. (2019).
\newblock {[Special Issue]Statistical Inference in the 21st Century: A World
  Beyond p < 0.05}.
\newblock {\em The American Statistician}, 73(1).

\bibitem[Wilks, 2018]{wilks2018enforcing}
Wilks, D.~S. (2018).
\newblock Enforcing calibration in ensemble postprocessing.
\newblock {\em Quarterly Journal of the Royal Meteorological Society},
  144(710):76--84.

\bibitem[Wilson et~al., 2007]{wilson2007calibrated}
Wilson, L.~J., Beauregard, S., Raftery, A.~E., and Verret, R. (2007).
\newblock {Calibrated surface temperature forecasts from the Canadian ensemble
  prediction system using Bayesian model averaging}.
\newblock {\em Monthly Weather Review}, 135(4):1364--1385.

\bibitem[Winkler et~al., 1996]{winkler1996scoring}
Winkler, R., Mu{\~n}oz, J., Cervera, J., Bernardo, J., Blattenberger, G.,
  Kadane, J., Lindley, D., Murphy, A., Oliver, R., and R{\'\i}os-Insua, D.
  (1996).
\newblock Scoring rules and the evaluation of probabilities.
\newblock {\em Test}, 5(1):1--60.

\bibitem[Zamo et~al., 2016]{zamo2016gridded}
Zamo, M., Bel, L., Mestre, O., and Stein, J. (2016).
\newblock Improved gridded windspeed forecasts by statistical post-processing
  of numerical models with block regression.
\newblock {\em Weather and Forecasting}, 31(6):1929--1945.

\bibitem[Zamo et~al., 2014]{zamo2014benchmarkII}
Zamo, M., Mestre, O., Arbogast, P., and Pannekoucke, O. (2014).
\newblock {A benchmark of statistical regression methods for short-term
  forecasting of photovoltaic electricity production. Part II: Probabilistic
  forecast of daily production}.
\newblock {\em Solar Energy}, 105:804--816.

\bibitem[Zamo and Naveau, 2018]{zamo2018estimation}
Zamo, M. and Naveau, P. (2018).
\newblock {Estimation of the Continuous Ranked Probability Score with Limited
  Information and Applications to Ensemble Weather Forecasts}.
\newblock {\em Mathematical Geosciences}, 50(2):209--234.

\bibitem[Zhu et~al., 2002]{zhu2002economic}
Zhu, Y., Toth, Z., Wobus, R., Richardson, D., and Mylne, K. (2002).
\newblock The economic value of ensemble-based weather forecasts.
\newblock {\em Bulletin of the American Meteorological Society}, 83(1):73.

\end{thebibliography}

% \begin{biography}[example-image-1x1]{A.~One}
% Please check with the journal's author guidelines whether author biographies are required. They are usually only included for review-type articles, and typically require photos and brief biographies (up to 75 words) for each author.
% \bigskip
% \bigskip
% \end{biography}

% \graphicalabstract{example-image-1x1}{Please check the journal's author guildines for whether a graphical abstract, key points, new findings, or other items are required for display in the Table of Contents.}

\end{document}